\RequirePackage{fix-cm}
\documentclass[epjc3,twocolumn]{svjour3}
\usepackage{booktabs}
\usepackage{xurl}
\journalname{Eur. Phys. J. C}

\RequirePackage{subfig}
\RequirePackage{amsfonts}
\RequirePackage{mathtools}
\RequirePackage{lipsum}
\RequirePackage{float}
\RequirePackage{dcolumn}
\RequirePackage{bm}
\RequirePackage{latexsym,color}
\RequirePackage{dcolumn}

\RequirePackage{amsmath}
\RequirePackage{amssymb}
\RequirePackage[T1]{fontenc}

\RequirePackage{graphicx}
\RequirePackage{mathptmx}      
\RequirePackage{flushend}
\RequirePackage[numbers,sort&compress]{natbib}
\RequirePackage[colorlinks,citecolor=blue,urlcolor=blue,linkcolor=blue]{hyperref}

\usepackage{ulem}
\usepackage{makecell}

\RequirePackage{color}

\newcommand{\bea}{\begin{eqnarray}}
\newcommand{\eea}{\end{eqnarray}}
\newcommand{\be}{\begin{equation}}
\newcommand{\ee}{\end{equation}}

 \catcode`\@=11 \@addtoreset{equation}{section}\catcode`\@=12

\newcommand{\beq}[1]{\begin{equation}\label{#1}}
\newcommand{\eeq}{\end{equation}}

\newcommand{\bear}[1]{\begin{eqnarray}\label{#1}}
\newcommand{\bearr}[1]{\begin{eqnarray}\lal \label{#1}}
\newcommand{\ear}{\end{eqnarray}}

\renewcommand{\thesection}{\arabic{section}}

\catcode`\@=11 \@addtoreset{equation}{section}\catcode`\@=12

\newcommand{\hrho}{\hat\rho_0}
\newcommand{\eff}{{\rm eff}}
\newcommand{\Msun}{M_\odot}
\newcommand{\doi}[1]{\href{https://doi.org/#1}{\nolinkurl{doi:#1}}}

\begin{document}

\title{Quasiperiodic oscillations as probes of rotating black holes in exponential dark matter environments}
\titlerunning{QPOs around rotating black holes in exponential dark matter environments}

\author{
M.~Alimkulova\thanksref{e1,addr1}
\and
K.~Boshkayev\thanksref{e2,addr1,addr2}
\and
Sh.~Gumarova\thanksref{e3,addr1}
\and
A.~Taukenova\thanksref{e4,addr1}
\and
A.~Urazalina\thanksref{e5,addr1}
}
\thankstext{e1}{e-mail: m.alimkulova@mail.ru}
\thankstext{e2}{e-mail: kuantay@mail.ru}
\thankstext{e3}{e-mail: sholpan5619@mail.ru}
\thankstext{e4}{e-mail: aliya\_tauken@mail.ru}
\thankstext{e5}{e-mail: y.a.a.707@mail.ru}
\institute{
Al-Farabi Kazakh National University, Al-Farabi Ave., 71, 050040, Almaty, Kazakhstan\label{addr1}
\and
 Kazakh‐-British Technical University, Tole bi str.~59, Almaty, 050000, Kazakhstan\label{addr2}
}
\date{Received: date / Accepted: date}
\maketitle
\begin{abstract}
We construct the rotating counterpart of a black hole embedded in a dark-matter halo with the exponential-sphere profile $\rho=\rho_0e^{-r/r_0}$, using the Newman--Janis algorithm in the Azreg-A\"{\i}nou form, and study its imprints on test-particle motion and quasiperiodic oscillations (QPOs). We derive closed-form orbital and epicyclic frequencies valid for any Kerr-like metric with a mass function and show that, to first order, the halo enters them only through its central density, acting as an effective cosmological constant $\Lambda_\eff=8\pi\rho_0$; the corrections are therefore universal for all cored profiles. The halo moves the innermost stable circular orbit outward and produces an outer boundary of stable orbits, while the sign of the frequency shift depends on the assumed equation of state of the halo. Fitting the relativistic precession model to the simultaneous QPO triplets of GRO~J1655$-$40 and XTE~J1859+226 with Markov-chain Monte Carlo sampling yields spins and emission radii consistent with Kerr and $\rho_0M^2<2.9\times10^{-5}$ and $3.0\times10^{-5}$ (95\%), respectively. In physical units these limits exclude only ultra-dense dark cores, $\rho_0\gtrsim10^{11}$~g\,cm$^{-3}$.

\end{abstract}

\section{Introduction}
\label{sec:intro}

Black holes are among the most profound predictions of Einstein's general relativity (GR) and serve as natural laboratories for probing gravity in its strongest regime. Although GR has been thoroughly verified in the weak-field domain through solar-system experiments \cite{Will2014}, and its strong-field predictions are being tested by the detection of gravitational waves \cite{LIGO2016} and by the direct imaging of the supermassive black holes M87$^*$ and Sgr~A$^*$ by the Event Horizon Telescope (EHT) \cite{EHT2019,EHT2022}, astrophysical black holes are never isolated. The dark-matter (DM) halos that dominate the mass budget of galaxies \cite{Bertone2005,NFW1996} extend down to the galactic nuclei, and the question of how a surrounding matter distribution modifies the near-horizon geometry has attracted considerable attention, both in the context of gravitational-wave astronomy \cite{Cardoso2022,Konoplya2022} and in the context of electromagnetic probes such as black-hole shadows \cite{Konoplya2019,Jusufi2019,Pantig2022}. A large body of work follows a common template: a phenomenological density profile is chosen, a static Schwarzschild-like metric is built from its mass function, the rotating counterpart is generated with the Newman--Janis algorithm (NJA) \cite{NewmanJanis1965,AzregAinou2014a,AzregAinou2014b}, and observables are computed. Profiles treated in this way include the Navarro--Frenk--White, Hernquist, Einasto, Burkert, Dehnen, King and Moore families \cite{Xu2018,Hou2018,Jusufi2023,AlBadawi2025,Uktamov2025,HeydariFard2026}.

One of the most intriguing challenges in modern astrophysics is to identify observable signatures capable of distinguishing black holes in modified geometries from those predicted by GR in the strong-field regime. The imprints of this regime are embedded in the radiation emitted by accretion flows: a substantial portion of this radiation originates deep inside the gravitational potential well and carries information about the near-horizon geometry \cite{Bardeen1972,Bambi2017book}. The motion of test particles is one of the most direct tools for constraining such geometries. In particular, the innermost stable circular orbit (ISCO) governs the structure of the accretion disc and the radiative efficiency, while small perturbations of circular orbits give rise to epicyclic frequencies that underpin the QPO phenomena observed in X-ray binaries \cite{StellaVietri1998,StellaVietri1999}.

QPOs are extracted from the X-ray light curves of accreting compact objects by Fourier analysis and are broadly divided into low-frequency (LF, $\lesssim30$~Hz) and high-frequency (HF, $\sim0.1$--1~kHz) types \cite{RemillardMcClintock2006,Belloni2012,IngramMotta2019,Motta2016}. HF QPOs are often detected in pairs whose frequency ratios in black-hole microquasars cluster around $3{:}2$ \cite{AbramowiczKluzniak2001,KluzniakAbramowicz2001,Torok2005,Strohmayer2001}. The most successful theoretical models link the QPO frequencies to the orbital and epicyclic frequencies of test particles near the ISCO; in the relativistic precession (RP) model \cite{StellaVietri1998,StellaVietri1999,2014GrCo...20..233B,2015ARep...59..441B,2024MNRAS.531.3876B, 2026MNRAS.549ag456B} the upper and lower HF QPOs and the type-C LF QPO are identified with the orbital, periastron-precession and nodal-precession frequencies of matter orbiting at a single radius, and the three simultaneous frequencies detected in GRO~J1655$-$40 fix the mass, the spin and the emission radius \cite{Motta2014a}. The sensitivity of these frequencies to the near-horizon geometry makes them suitable for testing alternative black-hole models \cite{Bambi2012,Bambi2015,Maselli2015,Allahyari2021,Jiang2021,Banerjee2022,Deligianni2021,Wang2022,Liu2023,2023PhRvD.108d4063B,2023PhRvD.108l4034B,Rayimbaev2023,Jumaniyozov2024,Shahzadi2021,Kolos2020,2025EPJC...85.1477B,Rahmatov2026,Khan2026,2026arXiv260909197M} and, more recently, black holes surrounded by DM halos \cite{Xamidov2025,Uktamov2025,Stuchlik2022,Ma2025}. Upcoming and current timing missions such as eXTP \cite{eXTP2019}, Insight-HXMT \cite{HXMT2020} and the Einstein Probe \cite{EP2025} are expected to yield more stringent constraints.

Recently the exponential-sphere profile,
\begin{equation}
\rho(r)=\rho_0\,e^{-r/r_0},
\label{eq:profile}
\end{equation}
introduced by Sofue to model the central mass distribution of the Milky Way \cite{Sofue2013}, has been used in numerous works \cite{2019MNRAS.484.3325B, 2020MNRAS.496.1115B,2021MNRAS.508.1543B,2022ApJ...925..210K, 2022ApJ...936...96B, 2024IJMPD..3350016B, 2026arXiv260423353S,2026arXiv260720262K, Boshkayev2026, 2026arXiv260724144A} to model various astrophysical scenario, including rotation curves of galaxies, the construction of a static Schwarzschild-like black hole submerged in a DM environment, whose shadow, photon sphere, quasinormal spectrum and gravitational wave signatures of periodic orbits were analyzed. The rotating generalization and its timing signatures have, to our knowledge, not been considered. In this work we construct the rotating counterpart through the NJA and perform a detailed analysis of test-particle dynamics and QPO phenomenology. Beyond this standard analysis we address three questions that apply to the whole class of ``black hole in a halo'' spacetimes: what a QPO actually measures on scales $r\ll r_0$, which way the frequencies move under different choices of the lapse function, and whether the resulting bounds are physically meaningful. The predictions are confronted with QPO data of microquasars through Bayesian inference.

The paper is organized as follows. Section~\ref{sec:metric} constructs the rotating spacetime and discusses its horizons, effective source and energy conditions. Section~\ref{sec:motion} is devoted to the motion of massive particles and the ISCO. Section~\ref{sec:freq} derives the fundamental frequencies and their universal first-order corrections. Section~\ref{sec:qpo} discusses twin-peak QPO models, Sect.~\ref{sec:mcmc} presents the Bayesian parameter constraints and their physical interpretation, and Sect.~\ref{sec:concl} summarizes the conclusions. The posterior distributions are collected in Appendix~\ref{app:corner}. Throughout we use $G=c=1$; hatted quantities are measured in units of the black-hole mass $M$ ($\hrho=\rho_0M^2$, $\hat r_0=r_0/M$).

\section{Rotating black hole in an exponential dark matter halo}
\label{sec:metric}

\subsection{Static seed metric}

The mass enclosed within radius $r$ by the profile (\ref{eq:profile}), together with a central black hole of mass $M$, is
\begin{equation}
\begin{aligned}
m(r)&=M+M_0\Big[1-e^{-r/r_0}\Big(1+\frac{r}{r_0}+\frac{r^2}{2r_0^2}\Big)\Big],\\
M_0&=8\pi\rho_0r_0^3,
\end{aligned}
\label{eq:mass}
\end{equation}
where $M_0$ is the total halo mass, so that the spacetime is asymptotically flat with ADM mass $M+M_0$. The static, spherically symmetric seed is taken in the Schwarzschild-like form
\begin{equation}
ds^2=-f(r)dt^2+f(r)^{-1}dr^2+r^2(d\theta^2+\sin^2\theta\,d\phi^2),
\label{eq:static}
\end{equation}
with the lapse function
\begin{equation}
f(r)=1-\frac{2m(r)}{r}.
\label{eq:lapse}
\end{equation}
Equations (\ref{eq:mass})--(\ref{eq:lapse}) coincide exactly with the solution of Boshkayev et al.\ \cite{Boshkayev2026}, who adopt the same mass function and set $g_{tt}=-g_{rr}^{-1}=-f$ (their $\mathcal{A}=\mathcal{B}=1-2m/r$). The metric reduces to Schwarzschild for $\rho_0\to0$ and to a regular, horizonless halo configuration for $M\to0$. Expanding for $r\ll r_0$,
\begin{equation}
m(r)=M+\varepsilon r^3\Big[1-\frac34\frac{r}{r_0}+\frac{3}{10}\frac{r^2}{r_0^2}-\ldots\Big],\quad \varepsilon\equiv\frac{4\pi}{3}\rho_0,
\label{eq:expansion}
\end{equation}
i.e.\ on scales small compared with $r_0$ the black hole is surrounded by a uniform core of density $\rho_0$, and the scale $r_0$ appears only through corrections suppressed by $r/r_0$. This structural fact, shared by every profile with a finite central density (Einasto, Burkert, King, pseudo-isothermal, Dehnen with $\gamma=0$), will be used repeatedly below.

\subsection{Rotating counterpart}

To describe rotating configurations we employ the NJA in the modified form of Azreg-A\"{\i}nou \cite{AzregAinou2014a,AzregAinou2014b}, which avoids the complexification step and produces a metric that can be written consistently in Boyer--Lindquist coordinates. For seeds of the form (\ref{eq:static}) with $g_{tt}g_{rr}=-1$ the result is the Kerr-like metric of G\"urses--G\"ursey type \cite{GursesGursey1975,BambiModesto2013,Toshmatov2014}
\begin{align}
ds^2=&-\Big(1-\frac{2m(r)r}{\Sigma}\Big)dt^2-\frac{4a\,m(r)\,r\sin^2\theta}{\Sigma}dt\,d\phi\nonumber\\
&+\frac{\Sigma}{\Delta}dr^2+\Sigma d\theta^2+\frac{\mathcal{A}}{\Sigma}\sin^2\theta\,d\phi^2,
\label{eq:kerrlike}
\end{align}
with
\begin{equation}
\begin{aligned}
&\Sigma=r^2+a^2\cos^2\theta,\qquad \Delta=r^2-2m(r)r+a^2,\\
&\mathcal{A}=(r^2+a^2)^2-a^2\Delta\sin^2\theta,
\end{aligned}
\label{eq:SigmaDelta}
\end{equation}
and $m(r)$ given by (\ref{eq:mass}). The metric reduces to the static seed (\ref{eq:static}) for $a\to0$ and to the Kerr geometry for $\rho_0\to0$. We note that the NJA was developed within GR and that rotating metrics generated by NJA-type procedures need not be exact solutions of a given set of field equations with a prescribed matter content \cite{Brahma2021,AzregAinou2014b}; for the present class of seeds, however, the metric (\ref{eq:kerrlike}) is an exact solution of the Einstein equations with the imperfect-fluid source discussed in Sect.~\ref{sec:source}.

\begin{figure*}[t]
\centering
\includegraphics[width=0.97\textwidth]{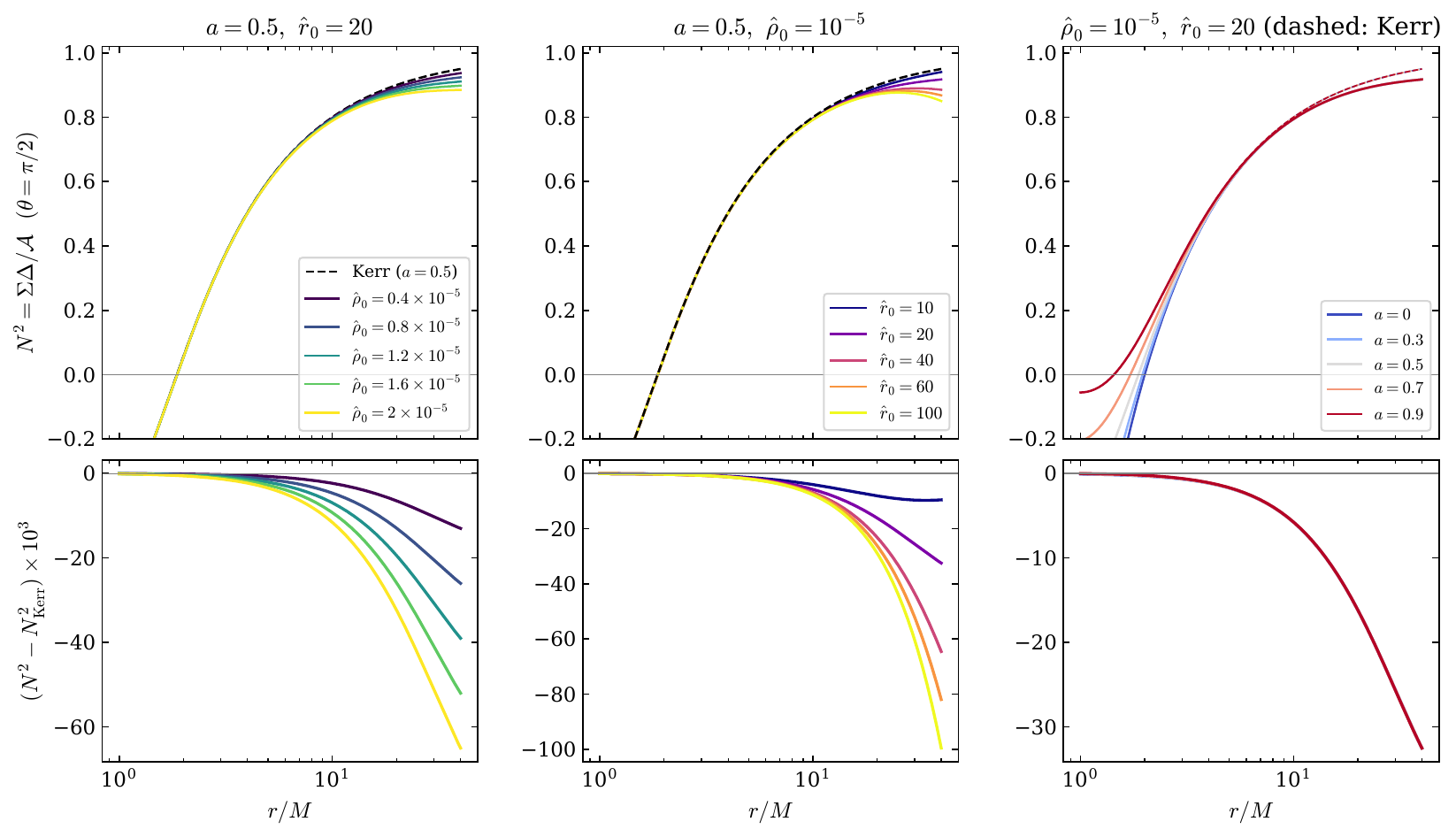}
\caption{Equatorial lapse function $N^2=\Sigma\Delta/\mathcal{A}$ of the rotating black hole in the exponential halo (upper row) and its deviation from the Kerr value with the same spin, $N^2-N^2_{\rm Kerr}$ (lower row, in units of $10^{-3}$). Left: variation of the core density $\hrho$ at $a=0.5$, $\hat r_0=20$; middle: variation of the halo scale $\hat r_0$ at $a=0.5$, $\hrho=10^{-5}$; right: variation of the spin at $\hrho=10^{-5}$, $\hat r_0=20$ (dashed curves: Kerr). The deviation is almost independent of the spin, so the five curves in the lower right panel overlap.}
\label{fig:lapse}
\end{figure*}

A convenient way to visualize the geometry is the lapse function of the zero-angular-momentum observers,
\begin{equation}
N^2=-\frac{1}{g^{tt}}=\frac{\Sigma\Delta}{\mathcal{A}},
\label{eq:N2}
\end{equation}
which reduces to $f(r)$ for $a=0$, vanishes on the horizons and measures the gravitational redshift of locally non-rotating observers. Figure~\ref{fig:lapse} shows $N^2$ in the equatorial plane together with its deviation from the Kerr value. The halo lowers $N^2$, i.e.\ deepens the gravitational potential, and the deviation grows outward, $N^2-N^2_{\rm Kerr}\simeq-2(m-M)/r\propto r^2$ for $r\ll r_0$, as expected from the enclosed halo mass; it is proportional to $\rho_0$, increases with $r_0$ (larger enclosed halo mass at fixed $\rho_0$) and is practically independent of the spin.

\begin{figure*}[t]
\centering
\includegraphics[width=2\columnwidth]{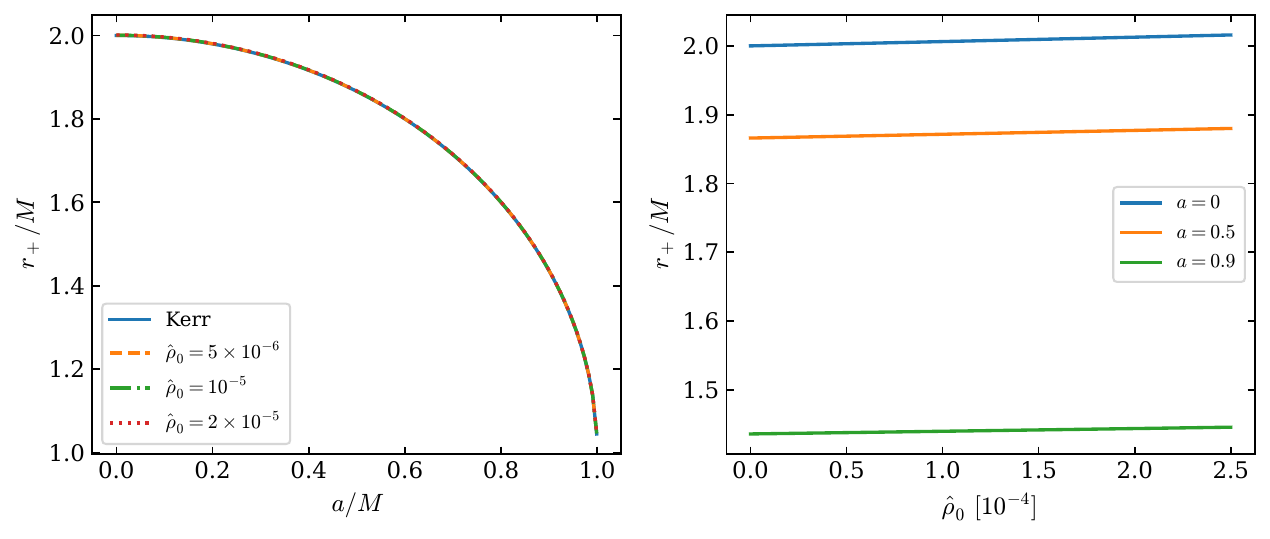}
\caption{Outer horizon radius as a function of the spin for several core densities (left) and as a function of the core density for several spins (right), for $\hat r_0=20$.}
\label{fig:horizon}
\end{figure*}

Because $m(r)$ is an increasing function, the outer horizon $r_+$, the largest root of $\Delta=0$, is slightly larger than its Kerr value (Fig.~\ref{fig:horizon}), $r_+\simeq M+\sqrt{M^2-a^2}+\varepsilon r_+^3+\ldots$, and the extremality bound is shifted above the Kerr value $a=M$ by a tiny amount, $a_{\max}\simeq m(r_+)\simeq M(1+\varepsilon M^2)$; numerically $\hat a_{\max}=1.00002$, $1.00004$ and $1.00008$ for $\hrho=5\times10^{-6}$, $10^{-5}$ and $2\times10^{-5}$ at $\hat r_0=20$. The ergoregion, bounded by $g_{tt}=0$, is likewise slightly enlarged.

If the halo is too compact, $\Delta$ acquires an additional pair of roots far from the black hole, a cosmological-like horizon. For the halo alone the condition $2m_h(r)/r<1$ reads $16\pi\rho_0r_0^2\,G(r/r_0)<1$ with $G(x)=[1-e^{-x}(1+x+x^2/2)]/x$, whose maximum is $G_{\max}=0.194$ at $x=3.38$, so that
\begin{equation}
\rho_0r_0^2<\frac{1}{16\pi G_{\max}}=0.102\ \Leftrightarrow\ \frac{M_0}{r_0}\lesssim2.6,
\label{eq:compact}
\end{equation}
a Buchdahl-type bound on the environment. For the fiducial scale $\hat r_0=20$ it gives $\hrho<2.6\times10^{-4}$; all parameter values used below satisfy it.

\subsection{Effective source and energy conditions}
\label{sec:source}

The effective matter source generating the geometry through the Einstein equations $G_{\mu\nu}=8\pi T_{\mu\nu}$ is obtained by projecting the stress--energy tensor onto the Carter orthonormal tetrad, $T_{\mu\nu}=\rho\,e^{(t)}_\mu e^{(t)}_\nu+p_re^{(r)}_\mu e^{(r)}_\nu+p_\theta e^{(\theta)}_\mu e^{(\theta)}_\nu+p_\phi e^{(\phi)}_\mu e^{(\phi)}_\nu$, with
$e^{(t)}=\sqrt{\Delta/\Sigma}\,(dt-a\sin^2\theta\,d\phi)$, $e^{(r)}=\sqrt{\Sigma/\Delta}\,dr$, $e^{(\theta)}=\sqrt\Sigma\,d\theta$, $e^{(\phi)}=(\sin\theta/\sqrt\Sigma)\,[(r^2+a^2)d\phi-a\,dt]$. For the class (\ref{eq:kerrlike}) one finds \cite{BambiModesto2013,Toshmatov2017,AzregAinou2014a}
\begin{equation}
\begin{aligned}
\rho&=-p_r=\frac{r^2m'}{4\pi\Sigma^2},\\
p_\theta&=p_\phi=-\frac{1}{8\pi}\Big[\frac{r\,m''}{\Sigma}+\frac{2a^2\cos^2\theta\,m'}{\Sigma^2}\Big],
\end{aligned}
\label{eq:source}
\end{equation}
where a prime denotes $d/dr$. We have verified Eq.~(\ref{eq:source}), including its $\theta$-dependence, by computing the Einstein tensor of the metric (\ref{eq:kerrlike}) symbolically for an arbitrary $m(r)$ and projecting it onto the tetrad: evaluated at randomly chosen points $(r,\theta,a)$ the residuals vanish to numerical precision ($\lesssim10^{-17}$), and the off-diagonal tetrad components are zero, so that the source is diagonal in the Carter frame. Using $\Sigma-r^2=a^2\cos^2\theta$, the angular pressures can be written equivalently as $p_\theta=p_\phi=\rho-(r\,m''+2m')/(8\pi\Sigma)$.

\begin{figure*}[t]
\centering
\includegraphics[width=0.97\textwidth]{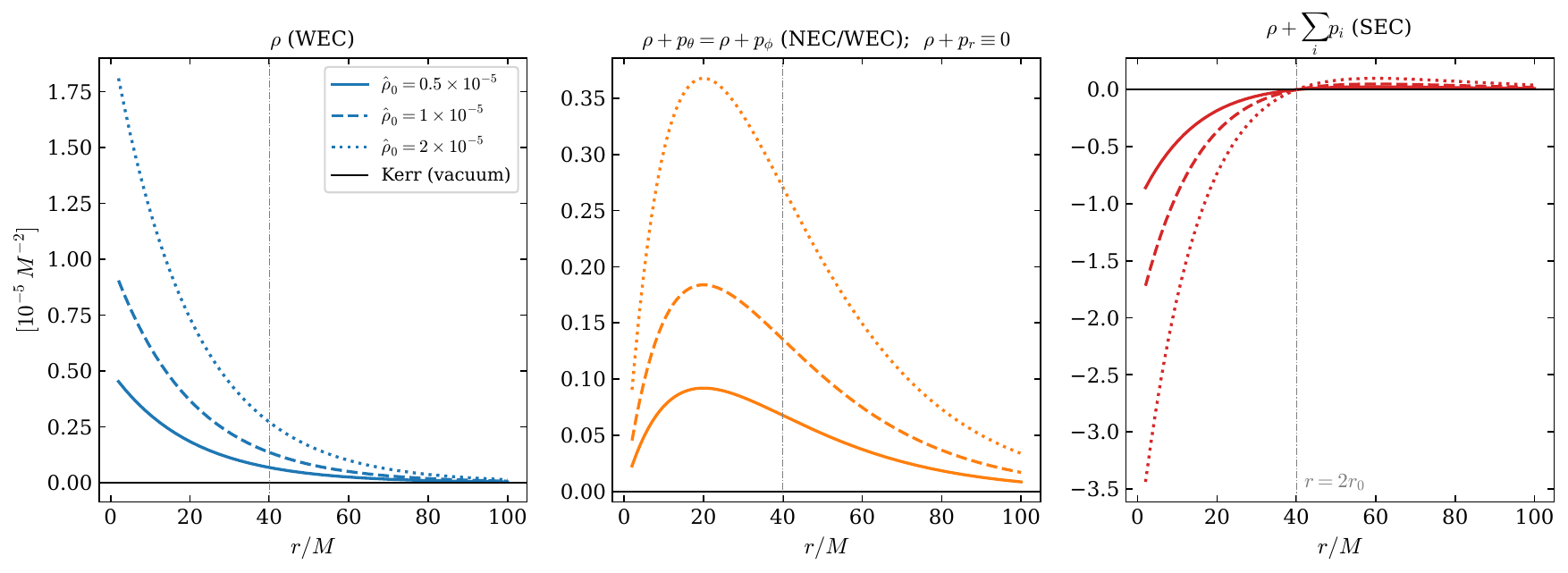}
\caption{Effective energy density and energy-condition combinations in the equatorial plane for $\hrho=5\times10^{-6}$ (solid), $10^{-5}$ (dashed) and $2\times10^{-5}$ (dotted), $\hat r_0=20$: $\rho$ (left), $\rho+p_\theta=\rho+p_\phi$ (middle; $\rho+p_r\equiv0$) and $\rho+\sum_ip_i$ (right). The black horizontal line is the Kerr (vacuum) value, zero for every combination. All curves scale linearly with $\hrho$; the strong energy condition is violated for $r<2r_0$ (vertical line). In the equatorial plane the source is independent of the spin.}
\label{fig:energy}
\end{figure*}

Since our analysis of particle motion and QPOs is restricted to the equatorial plane, we evaluate (\ref{eq:source}) at $\theta=\pi/2$, where $\Sigma=r^2$ and the source becomes independent of the spin,
\begin{equation}
\rho=\rho_0e^{-r/r_0},\quad p_r=-\rho,\quad p_\theta=p_\phi=-\rho\Big(1-\frac{r}{2r_0}\Big).
\label{eq:sourceeq}
\end{equation}
The weak and null energy conditions, $\rho\geq0$ and $\rho+p_i\geq0$, are satisfied ($\rho+p_r=0$, $\rho+p_\theta=r\rho/2r_0\geq0$), but the strong energy condition
\begin{equation}
\rho+\sum_ip_i=-2\rho\Big(1-\frac{r}{2r_0}\Big)<0\qquad(r<2r_0)
\label{eq:sec}
\end{equation}
is violated throughout the core. In the Kerr spacetime all these combinations vanish identically; Fig.~\ref{fig:energy} shows how they depart from zero for three values of the core density. The violation of the strong energy condition is not accidental: near the black hole, Eq.~(\ref{eq:expansion}) gives
\begin{equation}
f\simeq1-\frac{2M}{r}-\frac{\Lambda_\eff}{3}r^2,\qquad \Lambda_\eff=8\pi\rho_0=6\varepsilon,
\label{eq:lambda}
\end{equation}
i.e.\ under the ansatz $g_{tt}g_{rr}=-1$ the halo core acts exactly like a positive cosmological constant, and the effective source is an anisotropic fluid with $p_r=-\rho$ rather than a distribution of matter. To leading order in $r/r_0$ the metric (\ref{eq:kerrlike}) is therefore the NJA image of Schwarzschild--de~Sitter, i.e.\ the Kerr-like metric with $\Delta=r^2-2Mr+a^2-\Lambda_\eff r^4/3$ (not Carter's Kerr--de~Sitter solution, with which it coincides only in the static limit), and the classical results on circular orbits in Schwarzschild--de~Sitter \cite{StuchlikHledik1999,StuchlikSlany2004} can be taken over at $a=0$.

\subsection{On the choice of the lapse function}
\label{sec:lapsechoice}

\begin{figure*}[t]
\centering
\includegraphics[width=0.9\textwidth]{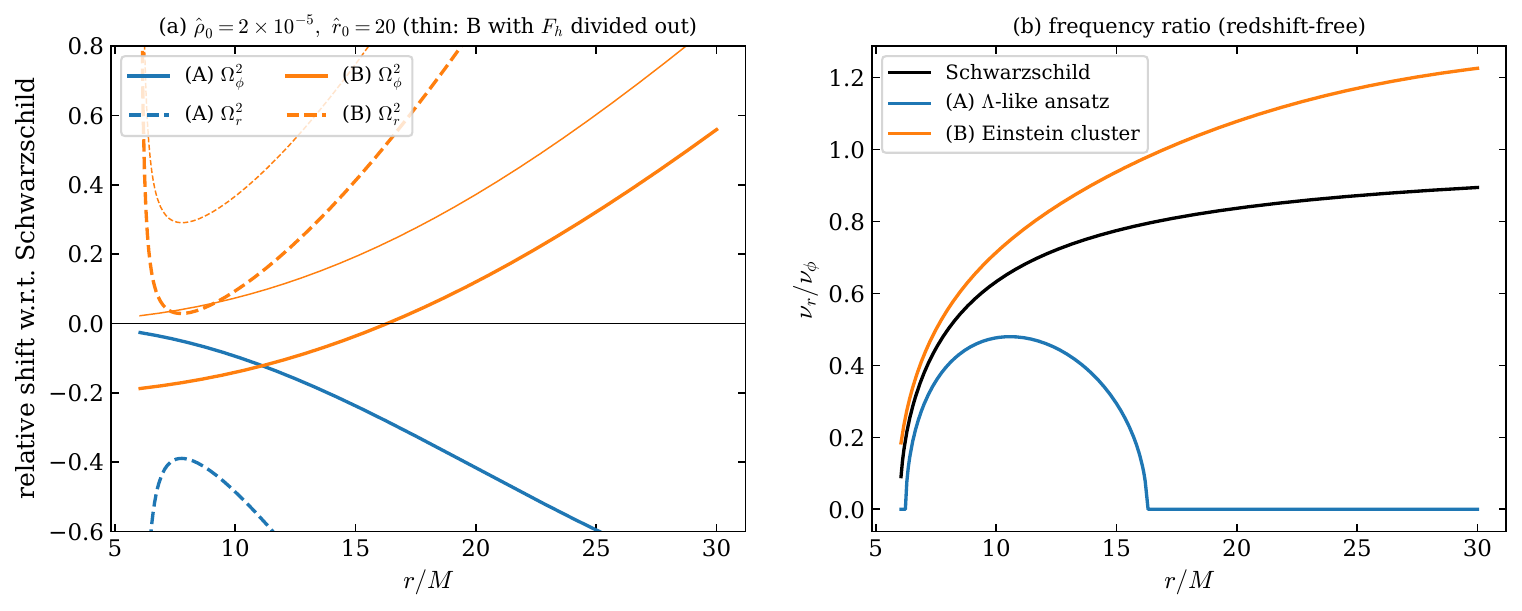}
\caption{Static comparison of the two constructions for $\hrho=2\times10^{-5}$, $\hat r_0=20$. Left: relative shifts of $\Omega_\phi^2$ (solid) and $\Omega_r^2$ (dashed) with respect to Schwarzschild for the $\Lambda$-like ansatz (A) and the Einstein cluster (B); thin lines show B with the halo lapse factor $F_h(r)$ divided out. Right: the redshift-free ratio $\nu_r/\nu_\phi$ moves in opposite directions in the two constructions.}
\label{fig:constructions}
\end{figure*}

The mass function fixes $g_{rr}=(1-2m/r)^{-1}$ through the $tt$ component of the Einstein equations, but the lapse $g_{tt}$ requires an equation of state for the environment. For an anisotropic fluid $T^\mu{}_\nu={\rm diag}(-\rho,p_r,p_t,p_t)$ one has
\begin{equation}
\frac{f'}{f}=\frac{2m+8\pi r^3p_r}{r(r-2m)}.
\label{eq:tov}
\end{equation}
The ansatz (\ref{eq:lapse}), used in most of the literature and hereafter called construction A, corresponds to $p_r=-\rho$. A halo of collisionless particles on circular orbits, the Einstein cluster \cite{Einstein1939}, has $p_r=0$ and
\begin{equation}
\frac{f_B'}{f_B}=\frac{2m}{r(r-2m)},\qquad p_t=\frac{\rho\,m}{2(r-2m)}>0,
\label{eq:cluster}
\end{equation}
which we call construction B; it is the construction used for the Hernquist halo in Ref.~\cite{Cardoso2022} and generalized in Ref.~\cite{Konoplya2022}, and it satisfies the strong energy condition. Both constructions share the same $g_{rr}$ and the same mass function; they differ only in $g_{tt}$ (see also the discussion of consistent solutions in Ref.~\cite{Bolokhov2026}). The distinction matters for QPOs. For a static metric $ds^2=-f\,dt^2+g^{-1}dr^2+r^2d\Omega^2$ the angular frequency of circular orbits $\Omega_\phi$ and the radial epicyclic frequency $\Omega_r$, i.e.\ the angular frequency of small radial oscillations about a circular orbit (denoted $\kappa$ in Newtonian galactic dynamics), are given by $\Omega_\phi^2=f'/2r$, $\Omega_\theta=\Omega_\phi$ and
\begin{equation}
\Omega_r^2=\frac{g}{2}\Big(f''+\frac{3f'}{r}-\frac{2f'^2}{f}\Big).
\label{eq:staticfreq}
\end{equation}
In the Newtonian regime they reduce to
\begin{align}
{\rm A:}&\quad \Omega_\phi^2\simeq\frac{M}{r^3}-\frac{8\pi}{3}\rho_0,\quad\ \ \Omega_r^2\simeq\frac{M}{r^3}-\frac{32\pi}{3}\rho_0,\nonumber\\
{\rm B:}&\quad \Omega_\phi^2\simeq F_h\Big[\frac{M}{r^3}+\frac{4\pi}{3}\rho_0\Big],\nonumber\\
&\quad \Omega_r^2\simeq F_h\Big[\frac{M}{r^3}+\frac{16\pi}{3}\rho_0\Big],
\label{eq:newton}
\end{align}
where $F_h(r)\equiv f_B(r)/(1-2M/r)$ is the halo part of the lapse, i.e.\ the gravitational redshift of the entire halo. In the Kerr/Schwarzschild limit $\Omega_r^2=\Omega_\phi^2(1-6M/r)$, so that $\Omega_r\to\Omega_\phi$ at large radii (closed Keplerian ellipses) and $\Omega_r\to0$ at the ISCO, where radial oscillations become unstable. The local corrections in (\ref{eq:newton}) have opposite signs and differ by a factor of two: construction A is repulsive, construction B attractive. In addition, construction B redshifts all frequencies by the slowly varying factor $F_h$, which for the toy parameters $\hrho=2\times10^{-5}$, $\hat r_0=20$ varies from $0.795$ at $r=6M$ to $0.808$ at $r=15M$ and is therefore almost exactly degenerate with the black-hole mass in any timing measurement. Figure~\ref{fig:constructions} illustrates both effects in full GR: at $r=8M$ construction A shifts $\Omega_\phi^2$ by $-5.4\%$ and $\Omega_r^2$ by $-39\%$, whereas construction B shifts them locally by $+4.3\%$ and $+29\%$. Frequency ratios such as $\nu_r/\nu_\phi$ are free of the global redshift and probe only the local core density. In the remainder of the paper we adopt construction A, which is the one underlying the static solution of Ref.~\cite{Boshkayev2026} and the rotating metric (\ref{eq:kerrlike}), keeping in mind that the direction of the induced parameter biases would be reversed for an attractive halo.

\section{Particle motion}
\label{sec:motion}

\begin{figure*}[t]
\centering
\includegraphics[width=2\columnwidth]{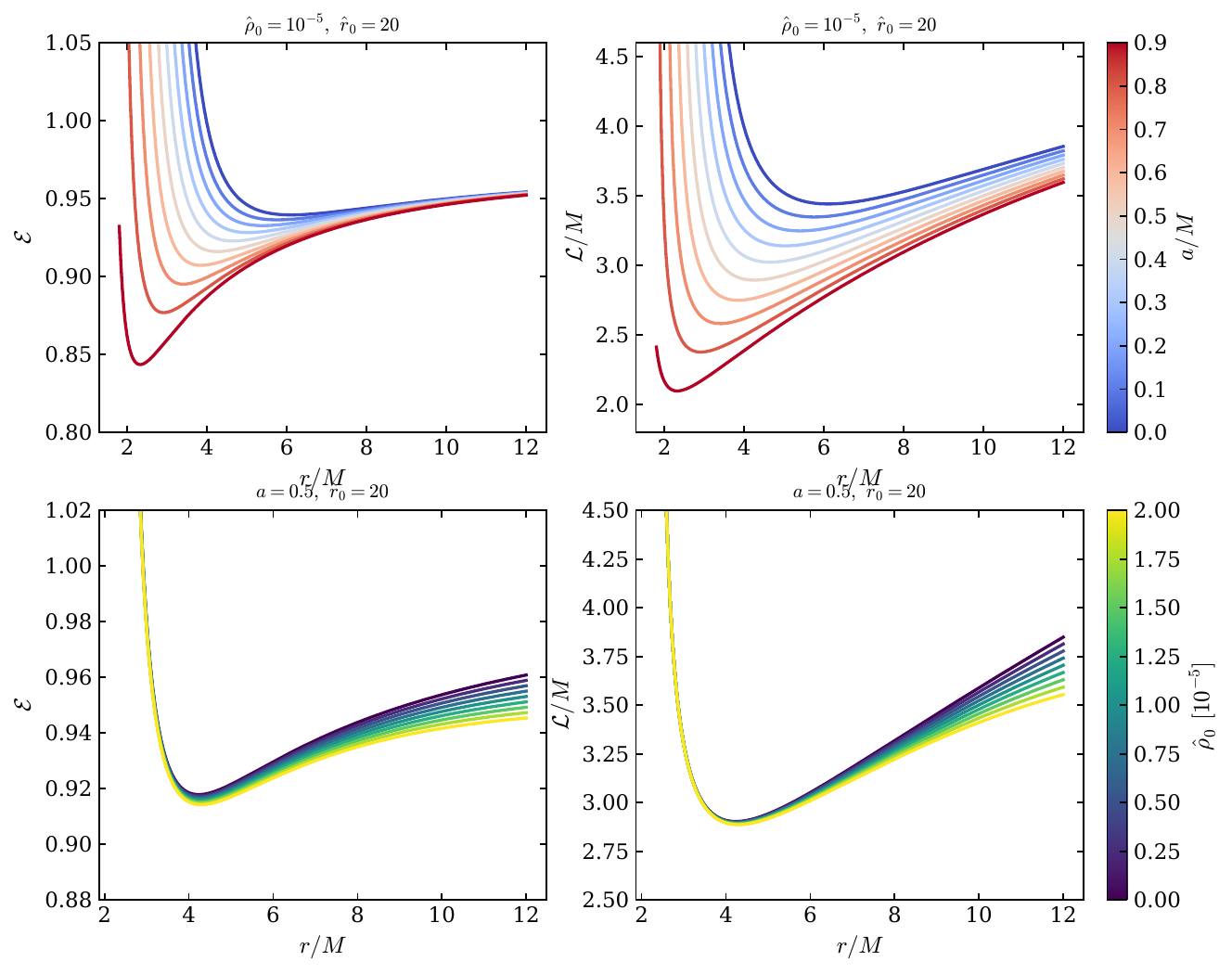}
\caption{Radial dependence of the specific energy (left) and angular momentum (right) of circular orbits for different values of the spin at fixed $\hrho=10^{-5}$ (upper panels) and of the core density at fixed $a=0.5$ (lower panels), with $\hat r_0=20$.}
\label{fig:EL}
\end{figure*}

The dynamics of a test particle of mass $\mu_p$ in the spacetime (\ref{eq:kerrlike}) is described by the Lagrangian $L_p=\frac12\mu_pg_{\mu\nu}\dot x^\mu\dot x^\nu$, where the dot denotes differentiation with respect to the affine parameter. Stationarity and axial symmetry provide two conserved quantities, the specific energy $\mathcal{E}$ and the specific angular momentum $\mathcal{L}$,
\begin{equation}
\mathcal{E}=-(g_{tt}\dot t+g_{t\phi}\dot\phi),\qquad \mathcal{L}=g_{t\phi}\dot t+g_{\phi\phi}\dot\phi,
\label{eq:EL}
\end{equation}
and the normalization $g_{\mu\nu}u^\mu u^\nu=-1$ provides an additional condition. Restricting the motion to the equatorial plane $\theta=\pi/2$ one obtains the radial equation
\begin{equation}
g_{rr}\dot r^2=V_\eff\equiv\frac{g_{\phi\phi}\mathcal{E}^2+2g_{t\phi}\mathcal{E}\mathcal{L}+g_{tt}\mathcal{L}^2}{g_{t\phi}^2-g_{tt}g_{\phi\phi}}-1.
\label{eq:veff}
\end{equation}
A circular orbit of radius $r_c$ requires $V_\eff(r_c)=\partial_rV_\eff(r_c)=0$, which determine $\mathcal{E}$ and $\mathcal{L}$,
\begin{equation}
\mathcal{E}=-\frac{g_{tt}+\Omega_\phi g_{t\phi}}{\sqrt{-\mathcal{N}}},\qquad \mathcal{L}=\frac{g_{t\phi}+\Omega_\phi g_{\phi\phi}}{\sqrt{-\mathcal{N}}},
\label{eq:circEL}
\end{equation}
with $\mathcal{N}=g_{tt}+2\Omega_\phi g_{t\phi}+\Omega_\phi^2g_{\phi\phi}$ evaluated at $r=r_c$ and $\Omega_\phi$ the orbital angular velocity given in Sect.~\ref{sec:freq}. (We reserve $r_0$ for the halo scale; $r_c$ denotes a generic circular-orbit radius and $r_e$ the QPO emission radius of Sect.~\ref{sec:mcmc}.) Figure~\ref{fig:EL} shows how $\mathcal{E}$ and $\mathcal{L}$ vary with the radial coordinate for different values of $a/M$ and $\hrho$. As the spin increases both quantities decrease and their minima, which define the ISCO, shift inward, as in the Kerr case. The halo core acts in the opposite direction: increasing $\hrho$ raises $\mathcal{E}$ and $\mathcal{L}$ at $r\gtrsim5M$ and shifts the ISCO outward, and the curves develop the characteristic upward bending at large $r$ that reflects the growth of the de~Sitter-like term (\ref{eq:lambda}).

\begin{figure*}[t]
\centering
\includegraphics[width=0.9\textwidth]{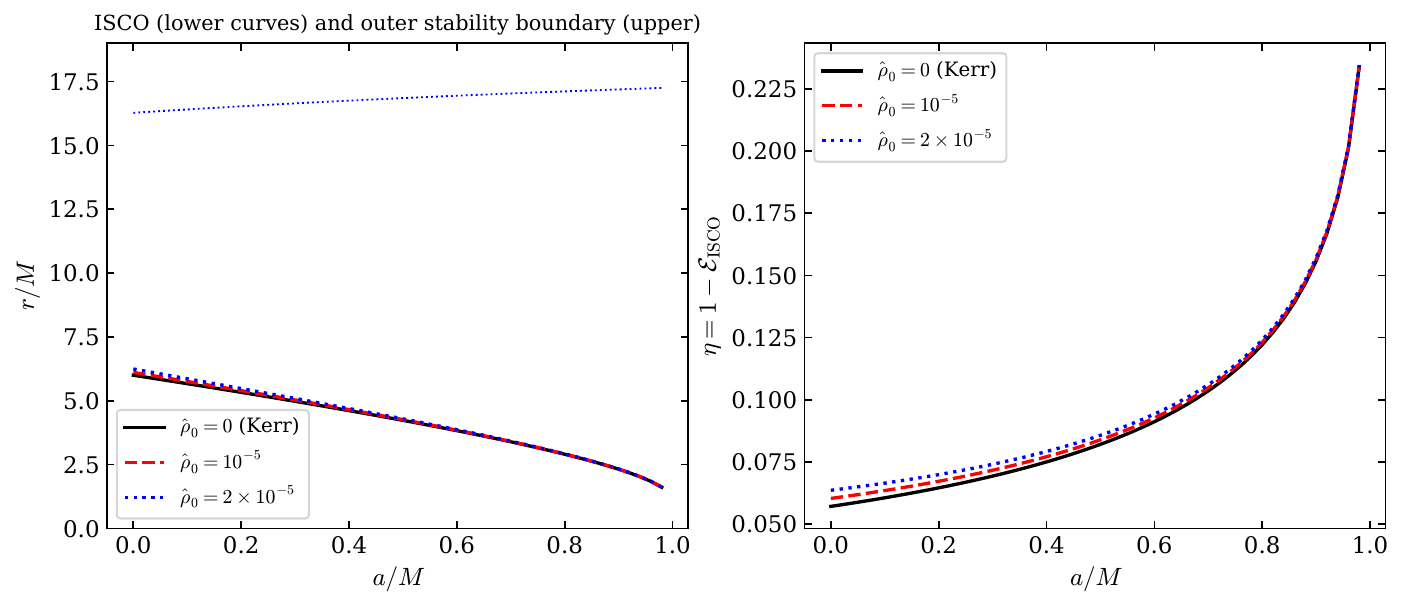}
\caption{Left: ISCO radius (lower curves) and outer boundary of stable circular orbits (upper curves, present only for $\hrho=2\times10^{-5}$ in this range) as functions of the spin. Right: accretion efficiency $\eta=1-\mathcal{E}_{\rm ISCO}$. Core densities $\hrho=0$ (Kerr), $10^{-5}$ and $2\times10^{-5}$; $\hat r_0=20$.}
\label{fig:isco}
\end{figure*}

The ISCO, where circular orbits change from stable to unstable, is obtained from $\partial_r^2V_\eff=0$, or equivalently from the vanishing of the radial epicyclic frequency. Figure~\ref{fig:isco} shows the ISCO radius and the accretion efficiency $\eta=1-\mathcal{E}_{\rm ISCO}$ as functions of the spin for several core densities. At $a=0$ the ISCO moves from $6.000M$ (Kerr) to $6.110M$ and $6.246M$ for $\hrho=10^{-5}$ and $2\times10^{-5}$, while at $a=0.9$ the shift is much smaller ($2.321M\to2.327M$), because the halo correction scales as $r^3$. The efficiency increases slightly with the core density, from $0.0572$ to $0.0604$ and $0.0637$ at $a=0$, since the repulsive de~Sitter-like term lowers the binding energy at the ISCO. A distinctive feature of the halo spacetime is the appearance of an outer boundary of the region of stable circular orbits (Fig.~\ref{fig:isco}), located at $r\simeq16.3M$ ($a=0$) and $17.2M$ ($a=0.9$) for $\hrho=2\times10^{-5}$: beyond it the growing halo term destabilizes radial oscillations, in complete analogy with the static radius of Schwarzschild--de~Sitter spacetimes \cite{StuchlikHledik1999}. The inner and outer boundaries merge, and stable circular orbits disappear altogether, above a spin-dependent critical core density,
\begin{equation}
\hat\rho_{\rm crit}(a)=5.1\times10^{-5},\ 8.2\times10^{-5},\ 1.2\times10^{-4},\ 2.2\times10^{-4}
\label{eq:rhocrit}
\end{equation}
for $a=0,\ 0.3,\ 0.5,\ 0.7$ ($\hat r_0=20$), while for $a\gtrsim0.9$ a narrow window of stable orbits at $r\simeq(2.3$--$4)M$ survives up to the compactness limit (\ref{eq:compact}). In the limit $r_0\to\infty$ (pure $r^3$ term) we find $\hat\rho_{\rm crit}(0)=2.86\times10^{-5}$, in agreement with the Schwarzschild--de~Sitter criterion $\Lambda M^2/3<12/15^4$ of Ref.~\cite{StuchlikHledik1999}, which gives $\hrho<2.83\times10^{-5}$; for finite $r_0$ the $O(r/r_0)$ corrections weaken the halo term at the merging radius $r\simeq(10$--$15)M$ and raise $\hat\rho_{\rm crit}$ accordingly.

\section{Fundamental frequencies}
\label{sec:freq}

\subsection{Orbital and epicyclic frequencies}

The orbital (Keplerian) angular velocity of circular equatorial orbits measured by a distant observer is
\begin{equation}
\Omega_\phi=\frac{d\phi}{dt}=\frac{-\partial_rg_{t\phi}+\sqrt{(\partial_rg_{t\phi})^2-\partial_rg_{tt}\,\partial_rg_{\phi\phi}}}{\partial_rg_{\phi\phi}}.
\label{eq:Omegaphi_gen}
\end{equation}
Substituting the metric coefficients of (\ref{eq:kerrlike}) we obtain the closed form
\begin{equation}
\Omega_\phi=\frac{\sqrt{\mu(r)}}{r^{3/2}+a\sqrt{\mu(r)}},\qquad \mu(r)\equiv m(r)-r\,m'(r),
\label{eq:Omegaphi}
\end{equation}
which reduces to the Kerr expression $\Omega_\phi=M^{1/2}/(r^{3/2}+aM^{1/2})$ for $m=M$ and shows that it is the local mass $m$ corrected by the local density ($rm'=4\pi r^3\rho$) that sets the orbital frequency. For the exponential halo $\mu(r)=M-2\varepsilon r^3[1-\frac98r/r_0+\ldots]$.

Small perturbations of a circular orbit, $r=r_c+\delta r$, $\theta=\pi/2+\delta\theta$, obey $\ddot{\delta r}+\Omega_r^2\delta r=0$ and $\ddot{\delta\theta}+\Omega_\theta^2\delta\theta=0$ with the radial and vertical epicyclic frequencies \cite{Ryan1995,Doneva2014}
\begin{align}
\Omega_r^2&=\frac{1}{2g_{rr}}\Big[X^2\partial_r^2\Big(\frac{g_{\phi\phi}}{D}\Big)-2XY\partial_r^2\Big(\frac{g_{t\phi}}{D}\Big)\nonumber\\
&\qquad\qquad+Y^2\partial_r^2\Big(\frac{g_{tt}}{D}\Big)\Big],\nonumber\\
\Omega_\theta^2&=\frac{1}{2g_{\theta\theta}}\Big[X^2\partial_\theta^2\Big(\frac{g_{\phi\phi}}{D}\Big)-2XY\partial_\theta^2\Big(\frac{g_{t\phi}}{D}\Big)\nonumber\\
&\qquad\qquad+Y^2\partial_\theta^2\Big(\frac{g_{tt}}{D}\Big)\Big],
\label{eq:epi_gen}
\end{align}
evaluated at $r=r_c$, $\theta=\pi/2$, where $D=g_{tt}g_{\phi\phi}-g_{t\phi}^2$, $X=g_{tt}+g_{t\phi}\Omega_\phi$ and $Y=g_{t\phi}+g_{\phi\phi}\Omega_\phi$. For the metric (\ref{eq:kerrlike}) these expressions, derived symbolically with $m$, $m'$ and $m''$ kept as independent symbols, collapse after elimination of the radicals to the compact closed forms
\begin{align}
\frac{\Omega_r^2}{\Omega_\phi^2}&=1-\frac{6\mu}{r}+\frac{8a\sqrt\mu}{r^{3/2}}-\frac{3a^2}{r^2}-2m'-\frac{m''\Delta}{\mu},\label{eq:Rr}\\
\frac{\Omega_\theta^2}{\Omega_\phi^2}&=1-\frac{4a\sqrt\mu}{r^{3/2}}+\frac{3a^2}{r^2}+\frac{2a\,m'}{\mu r}\big(a-2\sqrt{\mu r}\big),\label{eq:Rt}
\end{align}
which reduce to the Kerr ratios for $m'=m''=0$. We checked Eqs.~(\ref{eq:Omegaphi})--(\ref{eq:Rt}) against a direct numerical evaluation of (\ref{eq:Omegaphi_gen})--(\ref{eq:epi_gen}) for the exponential halo; they agree to machine precision ($|\Delta\Omega^2|\lesssim10^{-17}$). Equations~(\ref{eq:Omegaphi})--(\ref{eq:Rt}) hold for any Kerr-like metric of the form (\ref{eq:kerrlike}), i.e.\ for every rotating black hole generated by the NJA from a seed with $g_{tt}g_{rr}=-1$, and may be of independent use; note that the second derivative of the mass function enters only through $m''\Delta/\mu$, which vanishes on the horizon. In physical units the frequencies measured by a distant observer are
\begin{equation}
\nu_i=\frac{1}{2\pi}\frac{c^3}{GM}\,\Omega_i,\qquad i=\phi,r,\theta,
\label{eq:nuphys}
\end{equation}
i.e.\ $\nu_i\simeq3.23\times10^4\,\hat\Omega_i\,(M/\Msun)^{-1}$~Hz.

\begin{figure*}[t]
\centering
\includegraphics[width=0.97\textwidth]{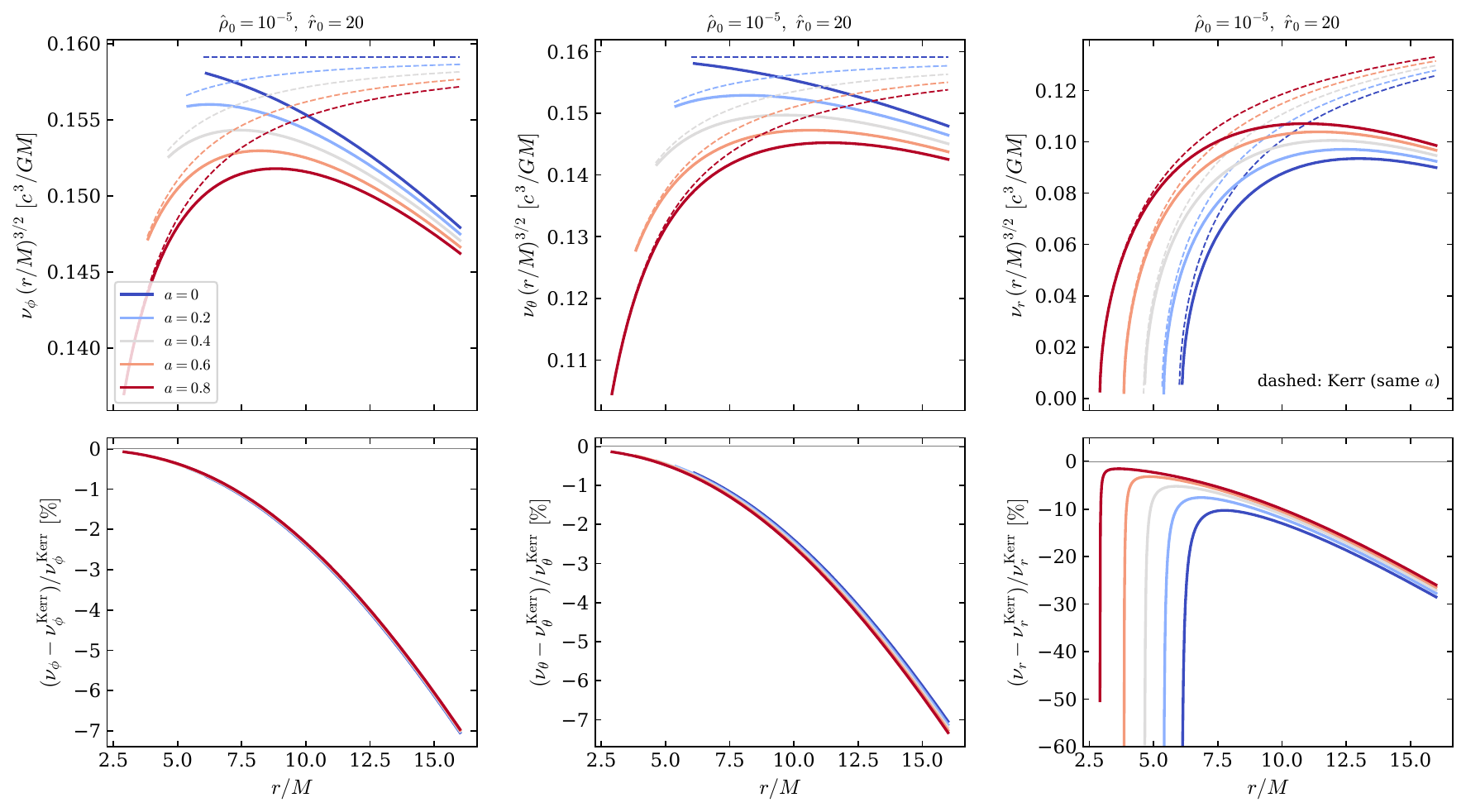}
\caption{Orbital ($\nu_\phi$, left), vertical ($\nu_\theta$, middle) and radial ($\nu_r$, right) epicyclic frequencies for several values of the spin, at fixed $\hrho=10^{-5}$, $\hat r_0=20$. Upper row: compensated frequencies $\nu_i(r/M)^{3/2}$ in units of $c^3/GM$ (solid: halo; dashed: Kerr with the same spin). Lower row: relative deviation from Kerr with the same spin, in per cent. Curves start at the ISCO of the respective spacetime.}
\label{fig:freqspin}
\end{figure*}

\begin{figure*}[t]
\centering
\includegraphics[width=0.97\textwidth]{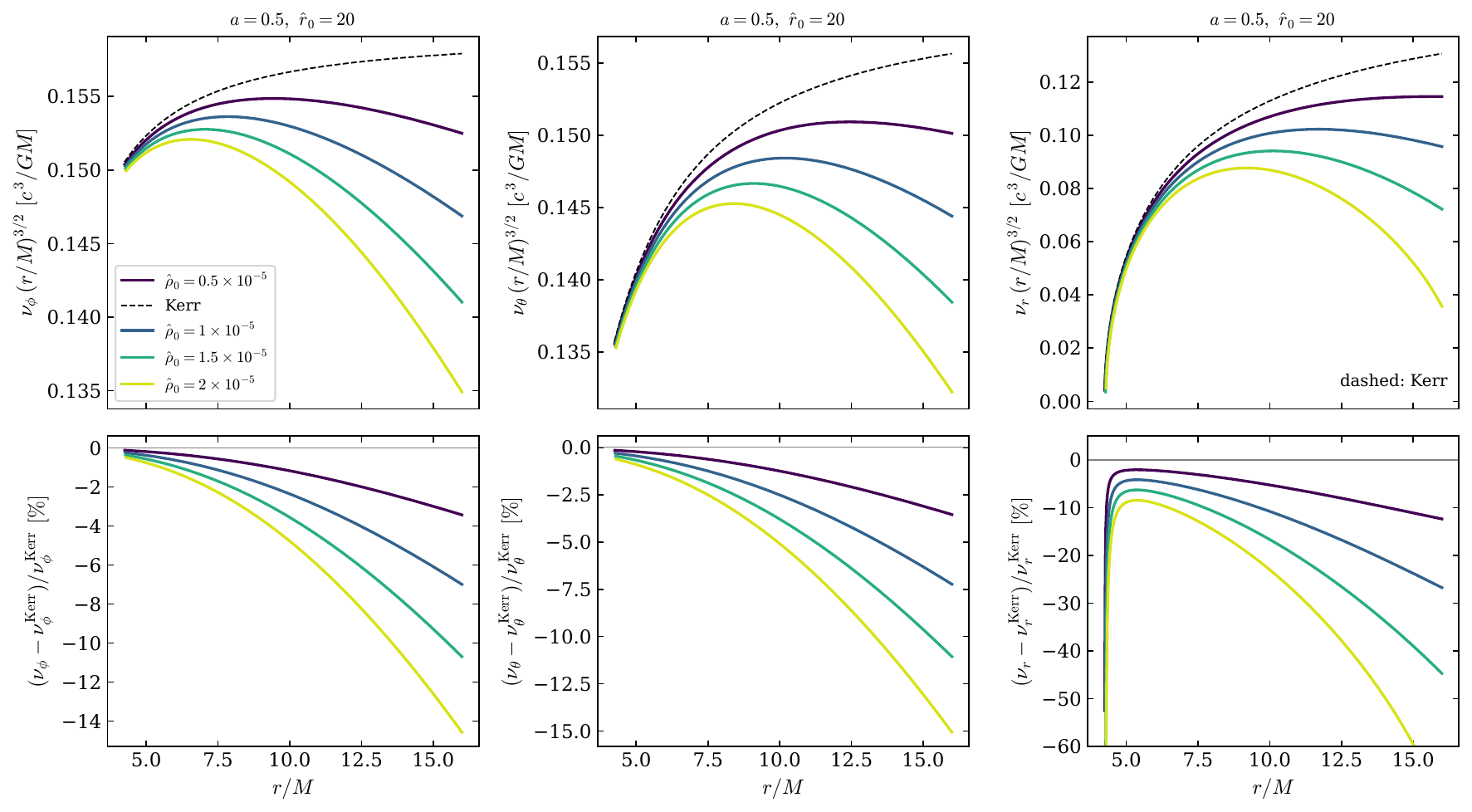}
\caption{Same as Fig.~\ref{fig:freqspin}, but for several values of the core density $\hrho$ at fixed spin $a=0.5$, $\hat r_0=20$ (dashed: Kerr, $a=0.5$).}
\label{fig:freqrho}
\end{figure*}

Figures~\ref{fig:freqspin} and \ref{fig:freqrho} illustrate the radial dependence of $\nu_\phi$, $\nu_\theta$ and $\nu_r$ for different values of the spin and of the core density. To make the differences visible the upper rows show the compensated frequencies $\nu_i(r/M)^{3/2}$, from which the Keplerian fall-off has been removed, and the lower rows show the relative deviation from the Kerr frequency with the same spin. For $a=0$ the orbital and vertical frequencies coincide, $\nu_\phi=\nu_\theta$, and as $a$ increases the ISCO moves inward, the peak of the radial epicyclic frequency shifts toward smaller radii and $\nu_\theta$ decreases more than $\nu_\phi$, as in the Kerr spacetime. The core density lowers all three frequencies at fixed radius and shifts the ISCO outward. For $\hrho=10^{-5}$ and $a=0.5$ the orbital frequency is reduced by $0.6\%$, $2.4\%$ and $6.1\%$ at $r=6M$, $10M$ and $15M$, the vertical frequency by $0.7\%$, $2.5\%$ and $6.3\%$, and the radial frequency by $4.4\%$, $11\%$ and $24\%$; at large radii $\nu_r$ is driven to zero at the outer stability boundary. The lower rows of Fig.~\ref{fig:freqspin} show that the relative corrections to $\nu_\phi$ and $\nu_\theta$ are practically independent of the spin (they differ by less than $0.4$ percentage points between $a=0$ and $0.9$ at $r\leq15M$), whereas the correction to $\nu_r$ is dominated by the spin-dependent position of the ISCO. The radial frequency responds about four times more strongly than the orbital one, as anticipated by the Newtonian limit (\ref{eq:newton}).

\subsection{Universal first-order corrections}

Inserting $m=M+\varepsilon r^3$ into (\ref{eq:Omegaphi})--(\ref{eq:Rt}) and expanding to first order in $\varepsilon$ around Kerr we find, in units $M=1$,
\begin{align}
\Omega_\phi^2&=\Omega_K^2\Big[1-2\varepsilon r^3\frac{r^{3/2}}{r^{3/2}+a}\Big],\label{eq:univ_phi}\\
\Omega_\theta^2&=\Omega_K^2\Big[\mathcal{K}_\theta-\tfrac{2\varepsilon r}{r^{3/2}+a}\big(r^{7/2}+4a^2r^{1/2}-3a^3\big)\Big],\label{eq:univ_th}\\
\Omega_r^2&=\Omega_K^2\Big[\mathcal{K}_r-\frac{2\varepsilon r}{r^{3/2}+a}\big(4r^{7/2}-15r^{5/2}+15ar^2\nonumber\\
&\qquad-9ar+4a^2r^{1/2}+3a^3\big)\Big],\label{eq:univ_r}
\end{align}
where $\Omega_K=(r^{3/2}+a)^{-1}$, $\mathcal{K}_r=1-6/r+8a/r^{3/2}-3a^2/r^2$ and $\mathcal{K}_\theta=1-4a/r^{3/2}+3a^2/r^2$ are the Kerr quantities. These formulae depend on the environment only through $\varepsilon=\frac{4\pi}{3}\rho_0$ and are therefore identical for every cored density profile; the profile-specific information enters at relative order $r/r_0$. In the static limit $\delta\Omega_\phi^2/\Omega_K^2=\delta\Omega_\theta^2/\Omega_K^2=-2\varepsilon r^3$ and $\delta\Omega_r^2/\Omega_K^2=-2\varepsilon r^2(4r-15)$, whose Newtonian limits are those of a de~Sitter term: the radial epicyclic frequency responds four times more strongly than the orbital one. The correction to $\Omega_r^2$ changes sign at $r=15M/4$, which explains why rapidly rotating black holes, whose ISCO lies inside this radius, retain a window of stable orbits at large core densities [cf.\ Eq.~(\ref{eq:rhocrit})]. The first-order ISCO shift follows from (\ref{eq:Rr}) as $\delta r_{\rm ISCO}\simeq3888\,\varepsilon_\eff M^3$ at $a=0$, with $\varepsilon_\eff=\varepsilon F(r/r_0)$ and $F$ defined in Eq.~(\ref{eq:rhoeff}) below; for $\hat r_0=20$, $\hrho=10^{-5}$ it gives $r_{\rm ISCO}=6.117M$ against the exact $6.110M$. We verified numerically that Eqs.~(\ref{eq:univ_phi})--(\ref{eq:univ_r}) reproduce the exact frequency shifts to better than a few per cent for $\hrho\lesssim10^{-5}$ at $r=6M$, the regime relevant for the observational constraints below.

\section{Quasi-periodic oscillations}
\label{sec:qpo}

\begin{figure*}[t]
\centering
\includegraphics[width=0.95\textwidth]{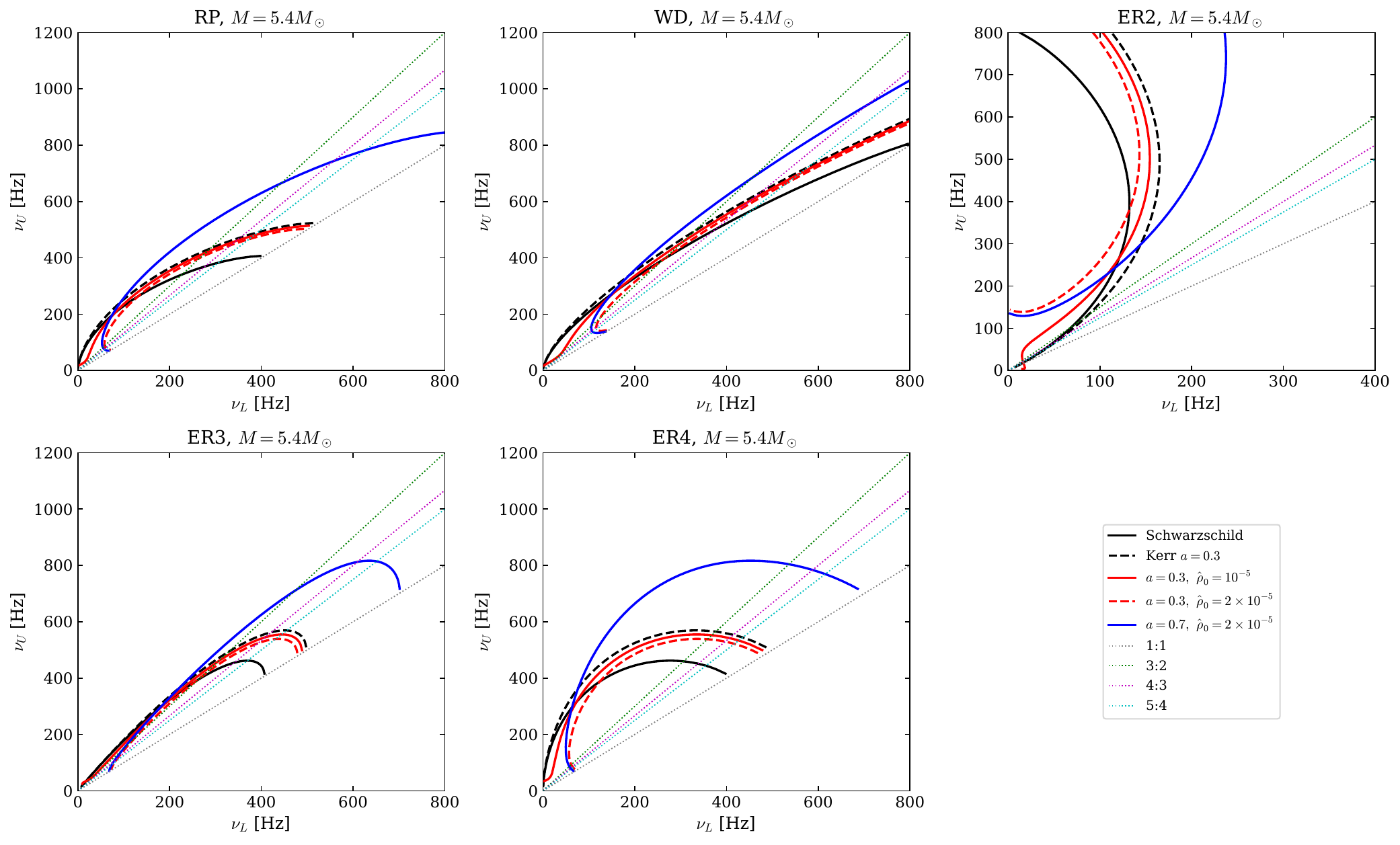}
\caption{Relations between the upper and lower frequencies of twin-peak QPOs in the RP, WD and ER2--ER4 models around the rotating black hole in the exponential halo, compared with the Schwarzschild and Kerr cases, for $M=5.4\,\Msun$ and $\hat r_0=20$. Dotted lines mark the frequency ratios $1{:}1$, $3{:}2$, $4{:}3$ and $5{:}4$.}
\label{fig:twin}
\end{figure*}

The predominant interpretation of the HF QPOs of black-hole microquasars relates them to quasi-harmonic oscillations of matter in the accretion disc at the fundamental frequencies of Sect.~\ref{sec:freq}. We consider the following standard twin-peak models \cite{Torok2005,Shahzadi2021,Kolos2020,Khan2026}: (I) the RP model \cite{StellaVietri1998,StellaVietri1999}, with $\nu_U=\nu_\phi$ and $\nu_L=\nu_\phi-\nu_r$; (II) the warped-disc (WD) model, with $\nu_U=2\nu_\phi-\nu_r$ and $\nu_L=2(\nu_\phi-\nu_r)$; (III) the epicyclic-resonance models ER2, ER3 and ER4, with $(\nu_U,\nu_L)=(2\nu_\theta-\nu_r,\nu_r)$, $(\nu_\theta+\nu_r,\nu_\theta)$ and $(\nu_\theta+\nu_r,\nu_\theta-\nu_r)$, respectively. Figure~\ref{fig:twin} shows the upper and lower frequencies produced along the sequence of circular orbits from the ISCO outward in the $\nu_U$--$\nu_L$ diagram, for a stellar-mass black hole with $M=5.4\,\Msun$ (GRO~J1655$-$40), together with the lines of constant ratio $3{:}2$, $4{:}3$ and $5{:}4$ and the $1{:}1$ line, the ``graveyard'' of twin-peak QPOs on which the two peaks merge into one.

Two features distinguish the halo spacetime from Kerr. First, in all models the halo lowers the frequencies at a given radius, so that the curves are shifted toward the lower-left part of the diagram and the intersection with the $3{:}2$ line, which fixes the resonance radius, moves outward. Second, whereas the Kerr curves extend to $\nu_L,\nu_U\to0$ at large radii, the halo curves terminate at the outer stability boundary, where $\nu_r\to0$: in the RP and WD models the curve ends on the $1{:}1$ line, while in ER2 and ER4 it ends on the axis $\nu_L=0$ and in ER3 on the $1{:}1$ line. A twin-peak QPO with a ratio close to unity generated far from the black hole is thus a generic signature of an environment of this type, in contrast to the Kerr case where such ratios only occur asymptotically. The spin and the core density act in opposite directions on the frequencies, which anticipates the degeneracy found in the Bayesian analysis of the next section.

\section{Parameter constraints from X-ray observations of QPOs}
\label{sec:mcmc}

\subsection{Observational data and the fitting procedure}
\label{sec:data}

For the quantitative analysis we adopt the RP model, which is the most extensively applied to the HF QPOs of microquasars and which, when a type-C LF QPO is detected simultaneously with the HF pair, provides three frequencies from a single emission radius $r_e$,
\begin{equation}
\nu_U=\nu_\phi,\qquad \nu_{\rm per}=\nu_\phi-\nu_r,\qquad \nu_{\rm nod}=\nu_\phi-\nu_\theta,
\label{eq:rpm}
\end{equation}
where $\nu_{\rm per}$ (identified with the lower HF QPO $\nu_L$) and $\nu_{\rm nod}$ are the periastron- and nodal-precession frequencies \cite{StellaVietri1999}. The observational input is listed in Table~\ref{tab:data}. Two sources have a simultaneous triplet and an independent, dynamical mass measurement: GRO~J1655$-$40, with the frequencies of Motta et al.\ \cite{Motta2014a} and the dynamical mass of Beer \& Podsiadlowski \cite{BeerPodsiadlowski2002}, and XTE~J1859+226, with the frequencies of Motta et al.\ \cite{Motta2022} and the dynamical mass of Yanes-Rizo et al.\ \cite{YanesRizo2022}. Note that the value $M=7.85\pm0.46\,\Msun$ quoted for XTE~J1859+226 in Ref.~\cite{Motta2022} is itself derived from the RP model in the Kerr metric and therefore cannot be used as a prior when the metric is varied; we use the optical dynamical mass $7.8\pm1.9\,\Msun$ instead. For H1743$-$322 the HF pair at $240$ and $165$~Hz \cite{Homan2005,IngramMotta2014} is not accompanied by a simultaneous type-C QPO: the $9.44$~Hz value is the highest type-C frequency ever detected in this source, and no dynamical mass is available \cite{IngramMotta2014}. This source is therefore treated separately in Sect.~\ref{sec:h1743}.

\begin{table*}[t]
\centering
\caption{QPO frequencies and mass measurements of the X-ray binaries used in the analysis. For GRO~J1655$-$40 and XTE~J1859+226 the three frequencies were detected simultaneously; for H1743$-$322 $\nu_{\rm nod}$ is the highest type-C frequency observed and is used only as a limit (Sect.~\ref{sec:h1743}).}
\label{tab:data}
\small\setlength{\tabcolsep}{3.2pt}
\begin{tabular}{lccccl}
\toprule
Source & $\nu_U=\nu_\phi$ (Hz) & $\nu_L=\nu_{\rm per}$ (Hz) & $\nu_{\rm nod}$ (Hz) & Dynamical $M$ ($\Msun$) & Refs.\ (QPOs; mass) \\
\midrule
GRO~J1655$-$40 & $441\pm2$ & $298\pm4$ & $17.3\pm0.1$ & $5.4\pm0.3$ & \cite{Motta2014a}; \cite{BeerPodsiadlowski2002}\\
XTE~J1859+226 & $227.5^{+2.1}_{-2.4}$ & $128.6^{+1.6}_{-1.8}$ & $3.65\pm0.01$ & $7.8\pm1.9$ & \cite{Motta2022}; \cite{YanesRizo2022}\\
H1743$-$322 & $240\pm3$ & $165^{+9}_{-5}$ & $9.44\pm0.02$ (highest type-C) & --- & \cite{Homan2005,IngramMotta2014}; ---\\
\bottomrule
\end{tabular}
\end{table*}

We stress that we do \emph{not} fit a functional relation $\nu_U(\nu_L)$, or its inverse, to a set of twin-peak pairs, as is done when many HF pairs of a source are available or in the resonance models of Sect.~\ref{sec:qpo}. Instead, for each source the three simultaneously measured frequencies are fitted jointly. For a parameter vector $\Theta=\{M,a/M,r_e/M,\hrho\}$ the model vector $\boldsymbol{\nu}_{\rm th}(\Theta)=(\nu_\phi,\nu_{\rm per},\nu_{\rm nod})$ is computed at the single radius $r_e$ from the closed-form frequencies (\ref{eq:Omegaphi})--(\ref{eq:Rt}), converted to Hz with Eq.~(\ref{eq:nuphys}) and combined according to (\ref{eq:rpm}), and compared with the observed vector $\boldsymbol{\nu}_{\rm obs}$ through the Gaussian likelihood
\begin{equation}
\ln\mathcal{L}(\Theta)=-\frac12\sum_{x\in\{\phi,{\rm per},{\rm nod}\}}\frac{\big(\nu_{x,{\rm obs}}-\nu_{x,{\rm th}}(\Theta)\big)^2}{\sigma_{x,{\rm obs}}^2},
\label{eq:like}
\end{equation}
where for asymmetric observational errors the larger side is used. In the Kerr metric ($\hrho=0$) the three equations (\ref{eq:rpm}) determine $M$, $a$ and $r_e$ uniquely; in the halo spacetime the fourth parameter $\hrho$ makes the problem underdetermined, and the information that breaks the resulting degeneracy comes from the mass prior and from the requirement that $r_e$ be a stable circular orbit.

\subsection{Bayesian inference: sampler, priors and the halo scale}
\label{sec:sampler}

The posterior distribution follows from Bayes' theorem, $P(\Theta|D)\propto\mathcal{L}(D|\Theta)P(\Theta)$, where $D$ denotes the observed frequencies. We sample it with the affine-invariant ensemble Markov-chain Monte Carlo method of Goodman \& Weare \cite{GoodmanWeare2010} as implemented in the \texttt{emcee} package (version 3.1) \cite{emcee2013}. In this method an \emph{ensemble} of $N_w$ ``walkers'' explores the parameter space simultaneously; each walker is an individual Markov chain whose position is updated at every step by a ``stretch move'' along the line joining it to a randomly selected walker of the complementary half of the ensemble, so that the proposal automatically adapts to the scale and correlations of the posterior. We use $N_w=32$ walkers (eight times the dimension of the parameter space), initialized in a small ball around the Kerr solution with $\hrho$ drawn uniformly from $[0,1.5\times10^{-5}]$. For GRO~J1655$-$40 every walker is advanced by $2\times10^4$ steps, of which the first $4\times10^3$ are discarded as burn-in; for XTE~J1859+226, whose posterior is a narrow curved ridge with longer autocorrelation, we use $10^5$ steps and discard $2\times10^4$. Convergence is monitored through the integrated autocorrelation time $\tau$ of each parameter \cite{emcee2013} and the mean acceptance fraction. For GRO~J1655$-$40 we find $\tau\simeq64$--$68$ steps and an acceptance fraction of $0.56$, i.e.\ the retained chains are $\sim240\tau$ long and contain $\simeq7600$ effectively independent samples; for XTE~J1859+226 $\tau\simeq900$--$2000$ steps, an acceptance fraction of $0.31$ and $\simeq1300$ independent samples (the retained chains are $\simeq40\tau$ long). The retained samples of all walkers are thinned by $\tau/2$ and merged into the posterior sample. As a further check we repeated the XTE~J1859+226 analysis with a five times shorter chain; the 95\% limit on $\hrho$ changed by less than 1\%.

The priors are summarized in Table~\ref{tab:priors}. The mass has a Gaussian prior centered on the dynamical value, $P(M)\propto\exp[-(M-\bar M)^2/2\sigma_M^2]$ (restricted to $M>0$); the spin, the emission radius and the core density have uniform priors, $a/M\in[0,0.998]$, $r_e/M\in[2,12]$ and $\hrho\in[0,10^{-4}]$. In addition, the prior vanishes unless $r_e$ is a stable circular orbit outside the horizon, i.e.\ unless $\Delta(r_e)>0$, $\mu(r_e)>0$, $\Omega_r^2(r_e)>0$ and $\Omega_\theta^2(r_e)>0$, which places $r_e$ between the ISCO and the outer stability boundary. The upper limit of the $\hrho$ prior lies well above the posterior support of both sources and above the critical density (\ref{eq:rhocrit}) for the relevant spins, and below the compactness bound (\ref{eq:compact}) $2.6\times10^{-4}$, so that every sampled configuration is free of a cosmological-like horizon. For the Kerr reference fits $\hrho=0$ and the remaining priors are unchanged.

\begin{table}[t]
\centering
\caption{Priors adopted in the MCMC analysis. $\mathcal{N}(\bar M,\sigma_M)$ is a Gaussian, $\mathcal{U}[x_1,x_2]$ a uniform distribution. In all fits $r_e$ must in addition be a stable circular orbit ($\Omega_r^2>0$, $\Omega_\theta^2>0$) outside the horizon; $\hat r_0=20$ unless stated otherwise.}
\label{tab:priors}
\begin{tabular}{lcc}
\toprule
Parameter & GRO~J1655$-$40 & XTE~J1859+226 \\
\midrule
$M/\Msun$ & $\mathcal{N}(5.4,\,0.3)$ & $\mathcal{N}(7.8,\,1.9)$ \\
$a/M$ & $\mathcal{U}[0,\,0.998]$ & $\mathcal{U}[0,\,0.998]$ \\
$r_e/M$ & $\mathcal{U}[2,\,12]$ & $\mathcal{U}[2,\,12]$ \\
$\hrho$ & $\mathcal{U}[0,\,10^{-4}]$ & $\mathcal{U}[0,\,10^{-4}]$ \\
\midrule
\multicolumn{3}{l}{Universality test (Sect.~\ref{sec:universality}), GRO~J1655$-$40:}\\
\multicolumn{3}{l}{\quad$\hat\rho_\eff\in\mathcal{U}[0,\,5\times10^{-5}]$, $\hat r_0=20,\,30,\,40$}\\
\bottomrule
\end{tabular}
\end{table}

\paragraph{Choice of the halo scale.} The halo scale is held fixed at $\hat r_0=20$ in the main fits for three reasons. (i) QPO frequencies are essentially insensitive to $r_0$: by Eq.~(\ref{eq:expansion}) the halo enters the frequencies at the emission radius only through the effective core density $\hat\rho_\eff=\hrho F(r_e/r_0)$ [Eq.~(\ref{eq:rhoeff})], and $r_0$ appears only through the slowly varying factor $F$; a free $r_0$ would therefore be completely degenerate with $\rho_0$ and not constrained by the data. We demonstrate this explicitly in Sect.~\ref{sec:universality}, where the fit is repeated for $\hat r_0=20$, 30 and 40. (ii) The compactness bound (\ref{eq:compact}) ties the two halo parameters together: the core densities to which QPOs are sensitive, $\hrho\sim10^{-5}$, are compatible with an asymptotically flat, horizon-free halo only for $\hat r_0\lesssim100$, and the full prior range $\hrho\leq10^{-4}$ requires $\hat r_0\lesssim32$. The value $\hat r_0=20$ satisfies this condition with a margin while keeping $r_e/r_0\simeq0.3$--$0.4$, so that the finite-size corrections of the profile are retained rather than neglected. (iii) Physically, $\hat r_0=20$ corresponds to $r_0\simeq150$--$230$~km for the two sources (depending on the adopted mass), i.e.\ to a compact dark core of the type discussed in Sect.~\ref{sec:physical}; there is no independent observational information on $r_0$ for such cores. For galactic-scale halos ($r_0\sim$~kpc, $\hat r_0\sim10^{15}$ for a stellar-mass black hole) the density accessible to QPOs is many orders of magnitude below the sensitivity, as discussed in Sect.~\ref{sec:physical}.

\subsection{Kerr reference fits}
\label{sec:kerrfit}

For comparison we first fit the Kerr metric ($\hrho=0$). The results are given in Table~\ref{tab:kerr}. For both sources the three RP-model equations have an exact solution inside the prior range, and the posterior width is set by the frequency errors. Our GRO~J1655$-$40 result agrees with Motta et al.\ \cite{Motta2014a}, who report $M=5.31\pm0.07\,\Msun$ and $a=0.290\pm0.003$, within $1.4\sigma$ in the spin, and our XTE~J1859+226 result reproduces $M=7.85\pm0.46\,\Msun$, $a=0.149\pm0.005$ of Ref.~\cite{Motta2022}. The QPOs originate at $r_e=5.68M$ and $6.85M$, i.e.\ $0.65M$ and $1.34M$ outside the respective ISCO.

\begin{table*}[t]
\centering
\caption{Kerr ($\hrho=0$) fits of the RP model: posterior medians and 68\% credible intervals, ISCO radius for the median spin, and the posterior of the distance of the emission radius from the ISCO.}
\label{tab:kerr}
\begin{tabular}{lccccc}
\toprule
Source & $M$ ($\Msun$) & $a/M$ & $r_e/M$ & $r_{\rm ISCO}/M$ & $(r_e-r_{\rm ISCO})/M$ \\
\midrule
GRO~J1655$-$40 (this work) & $5.303\pm0.061$ & $0.2861\pm0.0029$ & $5.680\pm0.034$ & $5.03$ & $0.65\pm0.04$\\
\quad Motta et al.\ \cite{Motta2014a} & $5.31\pm0.07$ & $0.290\pm0.003$ & --- & --- & ---\\
XTE~J1859+226 (this work) & $7.86\pm0.18$ & $0.1490\pm0.0019$ & $6.848^{+0.072}_{-0.068}$ & $5.50$ & $1.34^{+0.08}_{-0.06}$\\
\quad Motta et al.\ \cite{Motta2022} & $7.85\pm0.46$ & $0.149\pm0.005$ & --- & --- & ---\\
\bottomrule
\end{tabular}
\end{table*}

\subsection{Results for the exponential halo}
\label{sec:haloresults}

The best-fit parameters of the halo model are summarized in Table~\ref{tab:halo}, and the full posterior distributions, with 68\%, 90\% and 95\% credible contours, are shown in Appendix~\ref{app:corner} (Figs.~\ref{fig:cornerGRO} and \ref{fig:cornerXTE}). The spins and the emission radii are close to their Kerr values, and the QPO-generating region remains close to the ISCO, $r_e-r_{\rm ISCO}\simeq0.7M$ for GRO~J1655$-$40 and $\simeq1.9M$ for XTE~J1859+226; in all of 400 randomly drawn posterior samples $r_e$ lies at least $0.56M$ and $1.2M$, respectively, outside the ISCO of the corresponding spacetime. The core density, by contrast, is only bounded from above. With three measured frequencies and four parameters the problem is degenerate along a one-dimensional family of exact fits: a larger core density lowers all dimensionless frequencies [Eqs.~(\ref{eq:univ_phi})--(\ref{eq:univ_r})], which the fit compensates mainly by a lower mass and, to a lesser extent, by a larger spin and emission radius. This produces the strong correlations visible in the corner plots. We find
\begin{equation}
\begin{aligned}
\hrho&<2.9\times10^{-5}\quad\text{(GRO~J1655$-$40)},\\
\hrho&<3.0\times10^{-5}\quad\text{(XTE~J1859+226)}
\end{aligned}
\label{eq:bounds}
\end{equation}
at 95\% credibility. The two limits are similar in value but have different origins. For GRO~J1655$-$40 exact fits exist up to $\hrho\simeq5.5\times10^{-5}$, but along the degeneracy the mass falls from $5.30\,\Msun$ ($\hrho=0$) to $4.74\,\Msun$ ($\hrho=3\times10^{-5}$), $2.2\sigma$ below the precise dynamical mass, which is what limits the admissible core density. For XTE~J1859+226 the dynamical mass is too uncertain to play this role, but the family of exact fits itself terminates at $\hrho\simeq3.0\times10^{-5}$: the larger emission radius, $r_e\simeq7M$, enhances the $r^3$ halo term, and for larger core densities the required orbit would lie beyond the outer boundary of stable circular orbits, so that no stable orbit reproduces the observed triplet exactly. Beyond this point the smallest attainable $\chi^2$ grows rapidly ($\chi^2_{\min}\simeq0.8$, $3.1$ and $6.8$ at $\hrho=3.1$, $3.2$ and $3.3\times10^{-5}$), so that the posterior of $\hrho$ for this source rises gently up to $\simeq3\times10^{-5}$ and is cut off sharply between $3.0$ and $3.3\times10^{-5}$ (Fig.~\ref{fig:cornerXTE}).

\begin{table*}[t]
\centering
\caption{RP-model fits for the rotating black hole in the exponential halo ($\hat r_0=20$): posterior medians and 68\% credible intervals, 95\% upper limit on the core density, ISCO radius for the median spin and core density, and the distance of the emission radius from the ISCO. The median of $\hrho$ is quoted only for completeness; both posteriors are consistent with $\hrho=0$.}
\label{tab:halo}
\small\setlength{\tabcolsep}{4.5pt}
\begin{tabular}{lccccccc}
\toprule
Source & $M$ ($\Msun$) & $a/M$ & $r_e/M$ & $\hrho$ [$10^{-5}$] & $\hrho^{95\%}$ [$10^{-5}$] & $r_{\rm ISCO}/M$ & $(r_e-r_{\rm ISCO})/M$ \\
\midrule
GRO~J1655$-$40 & $5.14^{+0.13}_{-0.20}$ & $0.290^{+0.006}_{-0.004}$ & $5.78^{+0.13}_{-0.08}$ & $1.06^{+1.09}_{-0.75}$ & $<2.92$ & $5.07$ & $0.70^{+0.09}_{-0.06}$\\
XTE~J1859+226 & $6.68^{+0.85}_{-1.27}$ & $0.164^{+0.024}_{-0.011}$ & $7.52^{+1.00}_{-0.50}$ & $2.00^{+0.86}_{-1.30}$ & $<3.02$ & $5.62$ & $1.91^{+1.07}_{-0.45}$\\
\bottomrule
\end{tabular}
\end{table*}

Because construction B has the opposite local sign (Sect.~\ref{sec:lapsechoice}), the biases on $M$ and $a$ would be reversed for an attractive halo, while the magnitude of the bound on $|\varepsilon|$, set by the precision of the frequencies and of the dynamical masses, would be essentially the same.

\subsection{H1743$-$322: lower limits without a dynamical mass}
\label{sec:h1743}

For H1743$-$322 only the HF pair is available together with an upper envelope of the type-C frequency, and no dynamical mass. Following Ingram \& Motta \cite{IngramMotta2014}, for a trial mass $M$ we solve $\nu_\phi=240$~Hz and $\nu_{\rm per}=165$~Hz for $a$ and $r_e$, compute the nodal frequency at the ISCO and require $\nu_{\rm nod}(r_{\rm ISCO})\geq9.44$~Hz, since no type-C QPO can be produced inside the ISCO. In the Kerr metric this yields $M\geq9.28\,\Msun$, $a\geq0.215$ and $r_e\leq5.89M$, reproducing the limits $M\geq9.29\,\Msun$, $a\geq0.21$, $r\leq5.89M$ of Ref.~\cite{IngramMotta2014} and validating our implementation. In the halo spacetime the same procedure gives $M\geq8.92\,\Msun$ ($a\geq0.213$, $r_e\leq6.02M$) for $\hrho=10^{-5}$ and $M\geq8.50\,\Msun$ ($a\geq0.211$, $r_e\leq6.19M$) for $\hrho=2\times10^{-5}$. The halo therefore weakens the lower mass limit, as expected from the mass--density degeneracy, but without an independent mass measurement it does not constrain $\hrho$; a future dynamical mass for H1743$-$322 would turn this degeneracy into a bound comparable to (\ref{eq:bounds}).

\subsection{Universality of the bound}
\label{sec:universality}

The frequencies at the emission radius depend on the halo through $\mu(r_e)$ and its derivatives, Eqs.~(\ref{eq:Omegaphi})--(\ref{eq:Rt}). To leading order this is a single number, the effective core density seen at $r_e$,
\begin{equation}
\begin{aligned}
\hat\rho_\eff&\equiv\hrho\,F(r_e/r_0),\\
F(x)&\equiv-\frac{\mu(r)-M}{2\varepsilon r^3}=1-\frac98x+\frac35x^2-\ldots,
\end{aligned}
\label{eq:rhoeff}
\end{equation}
with $F\to1$ for $r_0\to\infty$; for $r_e=5.78M$ one has $F=0.720$, $0.804$ and $0.849$ at $\hat r_0=20$, 30 and 40. If the universality argument of Sect.~\ref{sec:freq} is correct, the data should constrain $\hat\rho_\eff$ independently of $r_0$, whereas the inferred limit on $\hrho$ itself should scale approximately as $1/F$. To test this we repeated the analysis of GRO~J1655$-$40 for $\hat r_0=20$, 30 and 40, now with a uniform prior on $\hat\rho_\eff\in[0,5\times10^{-5}]$ and all other priors as in Table~\ref{tab:priors}. The prior range is wide enough that the posteriors vanish well inside it, and every sample satisfies the compactness bound (\ref{eq:compact}) ($\hrho\leq4.8\times10^{-5}$ in all chains). Larger halo scales were not used because the compactness bound, $\hrho<0.102/\hat r_0^2$, would then truncate the posterior (at $\hat r_0=60$ already at $\hrho\simeq2.8\times10^{-5}$).

\begin{figure*}[t]
\centering
\includegraphics[width=0.92\textwidth]{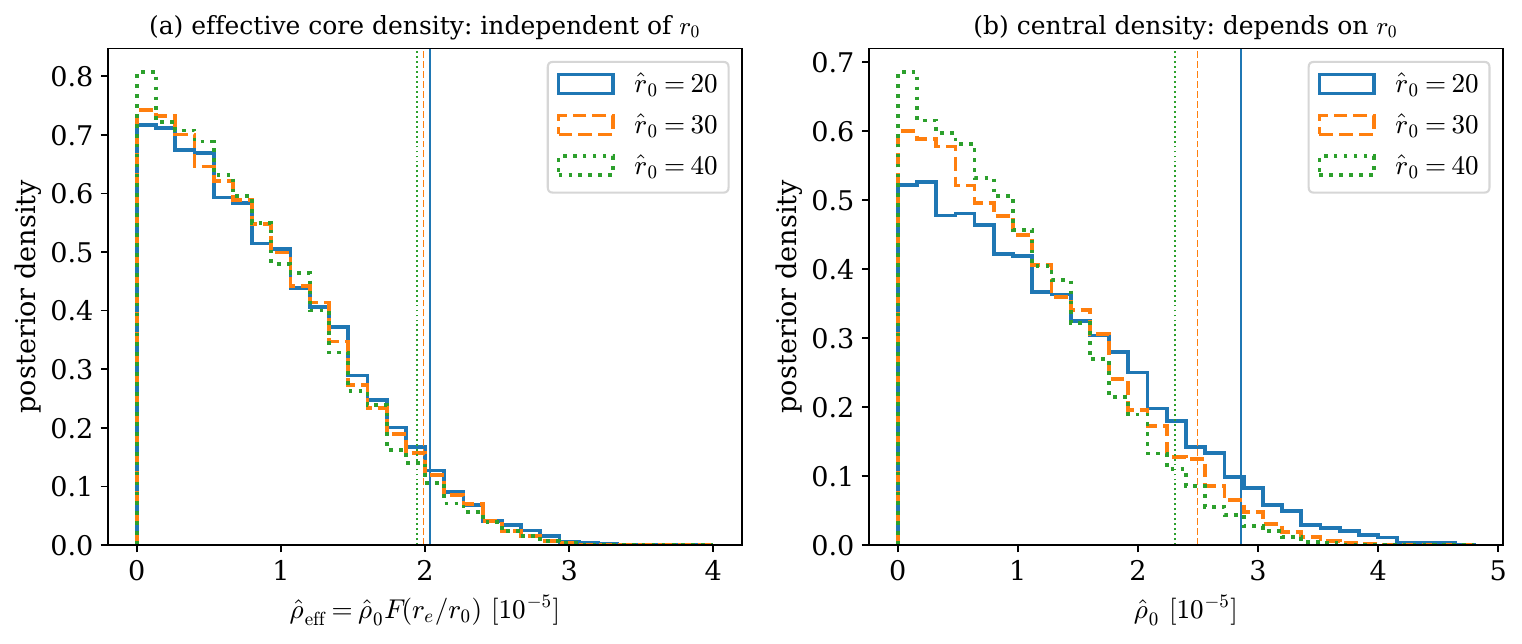}
\caption{Universality test for GRO~J1655$-$40. (a) Marginal posteriors of the effective core density $\hat\rho_\eff=\hrho F(r_e/r_0)$ from RP-model fits with $\hat r_0=20$ (solid), 30 (dashed) and 40 (dotted); they coincide within the sampling noise. (b) Marginal posteriors of the central density $\hrho$ obtained from the same chains, which depend on the assumed halo scale. Vertical lines mark the 95\% upper limits.}
\label{fig:universality}
\end{figure*}

Figure~\ref{fig:universality} shows the resulting marginal posteriors. Panel (a) displays the posterior of the effective core density $\hat\rho_\eff$ for the three halo scales: the three histograms coincide within the sampling noise, with 95\% limits $\hat\rho_\eff<2.03\times10^{-5}$, $1.99\times10^{-5}$ and $1.95\times10^{-5}$ (vertical lines), i.e.\ they agree to within 4\%. Panel (b) shows the posteriors of the central density $\hrho=\hat\rho_\eff/F$ obtained from the same chains: here the three distributions are visibly different, with 95\% limits $\hrho<2.86\times10^{-5}$, $2.50\times10^{-5}$ and $2.31\times10^{-5}$, which differ by 24\%, close to the ratio $F(\hat r_0=40)/F(\hat r_0=20)=1.18$ expected from the $1/F$ scaling. All posteriors decrease monotonically from zero, i.e.\ there is no preference for a non-vanishing halo. Three conclusions follow. First, QPO timing measures the effective density of the environment at the emission radius, not the profile parameters separately: changing $r_0$ by a factor of two leaves the information content of the data unchanged. Second, the bound is independent of the detailed shape of the profile, as anticipated by the universality of the first-order corrections (\ref{eq:univ_phi})--(\ref{eq:univ_r}); a bound on $\hrho$ alone, quoted without specifying $r_0$, is ambiguous at the level of tens of per cent. Third, because the compactness bound (\ref{eq:compact}) requires $\hat r_0\lesssim100$ for $\hrho\gtrsim10^{-5}$, the $O(r_e/r_0)$ corrections are never entirely negligible within this toy model, and the bound is best quoted in terms of $\hat\rho_\eff$. At $\hat r_0=20$ the limit $\hrho<2.86\times10^{-5}$ obtained here agrees with the value $2.92\times10^{-5}$ of Table~\ref{tab:halo}, which was obtained with a prior uniform in $\hrho$ rather than in $\hat\rho_\eff$.

\subsection{Physical interpretation and comparison with astrophysical bounds}
\label{sec:physical}

\paragraph{Physical units.} The dimensionless core density is converted to physical units through
\begin{equation}
\rho_0=\hrho\,\frac{c^6}{G^3M^2}=2.2\times10^{16}\,\hrho\Big(\frac{M}{5.3\,\Msun}\Big)^{-2}{\rm g\,cm^{-3}},
\label{eq:conversion}
\end{equation}
so that the bounds (\ref{eq:bounds}) correspond to $\rho_0\lesssim7\times10^{11}$ and $4\times10^{11}$~g\,cm$^{-3}$ (for the posterior median masses) for GRO~J1655$-$40 and XTE~J1859+226, with $r_0=20\,GM/c^2\simeq150$--$200$~km and a halo mass $M_0=8\pi\hrho\hat r_0^3M\lesssim6M$ contained within a few hundred kilometers. For supermassive black holes the conversion factor scales as $M^{-2}$: $\hrho\times3.4\times10^4$~g\,cm$^{-3}$ for Sgr~A$^*$ ($4.3\times10^6\,\Msun$) and $\hrho\times1.5\times10^{-2}$~g\,cm$^{-3}$ for M87$^*$ ($6.5\times10^9\,\Msun$).

\paragraph{Comparison with expected environments.} The central densities of cored galactic halos are $\rho_c\sim(0.01$--$1)\,\Msun\,{\rm pc}^{-3}\simeq10^{-24}$--$10^{-22}$~g\,cm$^{-3}$ \cite{Burkert1995,Salucci2019}. The adiabatic growth of a black hole can compress the surrounding DM into a steep ``spike'' with $\rho\propto r^{-7/3}$ \cite{GondoloSilk1999,Sadeghian2013}, which formally reaches much higher densities close to the hole. If the DM particles self-annihilate, however, the spike cannot be denser than the value at which the annihilation time equals the age $t_{\rm BH}$ of the black hole, because any excess would have been annihilated: the inner spike is flattened into a plateau of density $\rho_{\rm ann}=m_\chi/(\langle\sigma v\rangle t_{\rm BH})$ \cite{GondoloSilk1999}, numerically $\rho_{\rm ann}\simeq2\times10^{-14}(m_\chi/100~{\rm GeV})(3\times10^{-26}~{\rm cm^3\,s^{-1}}/\langle\sigma v\rangle)(10^{10}~{\rm yr}/t_{\rm BH})$~g\,cm$^{-3}$. Even if annihilation were negligible and the spike reached $\sim10^{-8}$~g\,cm$^{-3}$ near the horizon, it would remain about eight orders of magnitude below the density that QPOs could probe around Sgr~A$^*$ [Eq.~(\ref{eq:conversion}) with $\hrho\sim3\times10^{-5}$] and nearly twenty orders below the bounds (\ref{eq:bounds}) for the stellar-mass black holes; equivalently, the fractional frequency shift $\delta\nu/\nu\sim\hrho\hat r_e^3$ expected from a realistic environment is $\lesssim10^{-10}$, far beyond the reach of any planned timing mission. This conclusion is independent of the profile, since it follows from the universality of Eq.~(\ref{eq:expansion}), and applies to every study that fits geometric-unit halo parameters to QPO data. Two positive statements can nevertheless be made. First, the bounds (\ref{eq:bounds}) are bona fide constraints on hypothetical dense dark cores in the immediate vicinity of the black hole (for instance self-gravitating configurations of ultralight bosons or other exotic compact environments with $\rho\sim10^{11}$~g\,cm$^{-3}$ and size $\sim10^2$~km), and they are the first such constraints for a rotating black hole with an exponential environment. Second, for realistic halos the leading observable is not the local core term but the global redshift of construction B, $\delta\nu/\nu\simeq\Phi_{\rm halo}(0)=-M_0/2r_0\sim-10^{-6}$ for a Milky-Way-like halo, which is degenerate with the mass: QPO masses of black holes deep inside galactic potential wells are biased low by the same fraction.

\paragraph{Complementarity with shadow constraints.} The static analysis of Ref.~\cite{Boshkayev2026} constrained the halo parameters from the EHT measurements of the shadows of M87$^*$ and Sgr~A$^*$, which probe the null-geodesic structure near the photon sphere, whereas the QPO method probes the timelike geodesics near the ISCO. The two approaches are complementary, and the parameter region allowed by the present analysis is consistent with the one not excluded by the shadow observables; a combined analysis is left for future work.

\section{Conclusions}
\label{sec:concl}

In this work we have constructed the rotating counterpart of a black hole submerged in an exponential-sphere dark-matter environment and investigated the geodesic motion of test particles and the QPO phenomena in its background. The main findings are summarized as follows.
\begin{itemize}
\item {\it Spacetime structure.} The rotating metric (\ref{eq:kerrlike}) is of Kerr-like form with the mass function (\ref{eq:mass}). The halo lowers the lapse function, slightly enlarges the horizon and the ergoregion and shifts the extremality bound marginally above $a=M$. The environment must satisfy the compactness bound $\rho_0r_0^2<0.102$, otherwise a cosmological-like horizon appears. In the equatorial plane the effective source reduces exactly to $\rho=\rho_0e^{-r/r_0}$, $p_r=-\rho$, $p_t=-\rho(1-r/2r_0)$, which satisfies the weak but violates the strong energy condition: under the ansatz $g_{tt}g_{rr}=-1$ the halo core acts as an effective positive cosmological constant $\Lambda_\eff=8\pi\rho_0$.
\item {\it Sign of the effect.} The ansatz is repulsive and lowers the frequencies, whereas an Einstein-cluster halo with $p_r=0$ is attractive, raises them and additionally redshifts all frequencies by a mass-degenerate factor. The direction of the inferred parameter biases is thus a modelling choice, and frequency ratios are the redshift-free observables.
\item {\it Particle dynamics and ISCO.} The core density raises the specific energy and angular momentum of circular orbits, moves the ISCO outward and slightly increases the accretion efficiency, while the spin acts in the opposite direction. The halo produces an outer boundary of the region of stable circular orbits, and stable orbits disappear above the spin-dependent critical core density (\ref{eq:rhocrit}), whose $r_0\to\infty$ limit reproduces the Schwarzschild--de~Sitter criterion.
\item {\it Fundamental frequencies.} We obtained compact closed forms (\ref{eq:Omegaphi})--(\ref{eq:Rt}) for the orbital and epicyclic frequencies of any Kerr-like metric with a mass function, and universal first-order corrections (\ref{eq:univ_phi})--(\ref{eq:univ_r}) that depend on the environment only through its central density and therefore hold for every cored profile. The relative corrections to $\nu_\phi$ and $\nu_\theta$ are nearly spin-independent, and the radial epicyclic frequency responds about four times more strongly than the orbital one.
\item {\it QPO models.} In the RP, WD and ER2--ER4 models the halo shifts the twin-peak curves toward lower frequencies and moves the $3{:}2$ resonance radius outward; the curves terminate at the outer stability boundary, so that twin peaks with ratios close to unity generated far from the black hole are a generic signature of such environments.
\item {\it Bayesian parameter estimation.} Fitting the RP model jointly to the simultaneous QPO triplets of GRO~J1655$-$40 and XTE~J1859+226, with dynamical-mass priors, gives spins and emission radii close to their Kerr values, with the emission radius $0.7M$--$1.9M$ outside the ISCO, and upper limits $\hrho<2.9\times10^{-5}$ and $3.0\times10^{-5}$ at 95\% credibility. The first is set by the precise dynamical mass, the second by the disappearance of stable orbits that reproduce the triplet. For H1743$-$322, which lacks a simultaneous triplet and a dynamical mass, the halo only weakens the lower mass limit. The bound is independent of the halo scale once expressed through the effective core density (\ref{eq:rhoeff}).
\item {\it Physical interpretation.} In physical units the limits, $\rho_0\lesssim(4$--$7)\times10^{11}$~g\,cm$^{-3}$, exclude only exotic ultra-dense environments; realistic galactic halos affect QPO frequencies at the level of $10^{-10}$ locally and $10^{-6}$ through the global redshift. This conclusion is profile independent.
\end{itemize}

Natural extensions include the rotating Einstein-cluster halo via the general Azreg-A\"{\i}nou transformation, the application of resonance models to the $3{:}2$ twin-peak sources with Bayesian model comparison, and the extension of the universality argument to cusped profiles, for which the leading correction scales differently with $r$ and is, in principle, distinguishable.

\begin{acknowledgements}
Generative AI tools were used to assist with symbolic derivations, code and language editing; all results were verified by the author, who takes full responsibility for the content.
\end{acknowledgements}

\section*{Data availability statement}
The symbolic derivations, numerical routines and MCMC chains underlying this article are available from the corresponding author on reasonable request.

\appendix
\section{Posterior distributions}
\label{app:corner}

\begin{figure*}[p]
\centering
\includegraphics[width=0.85\textwidth]{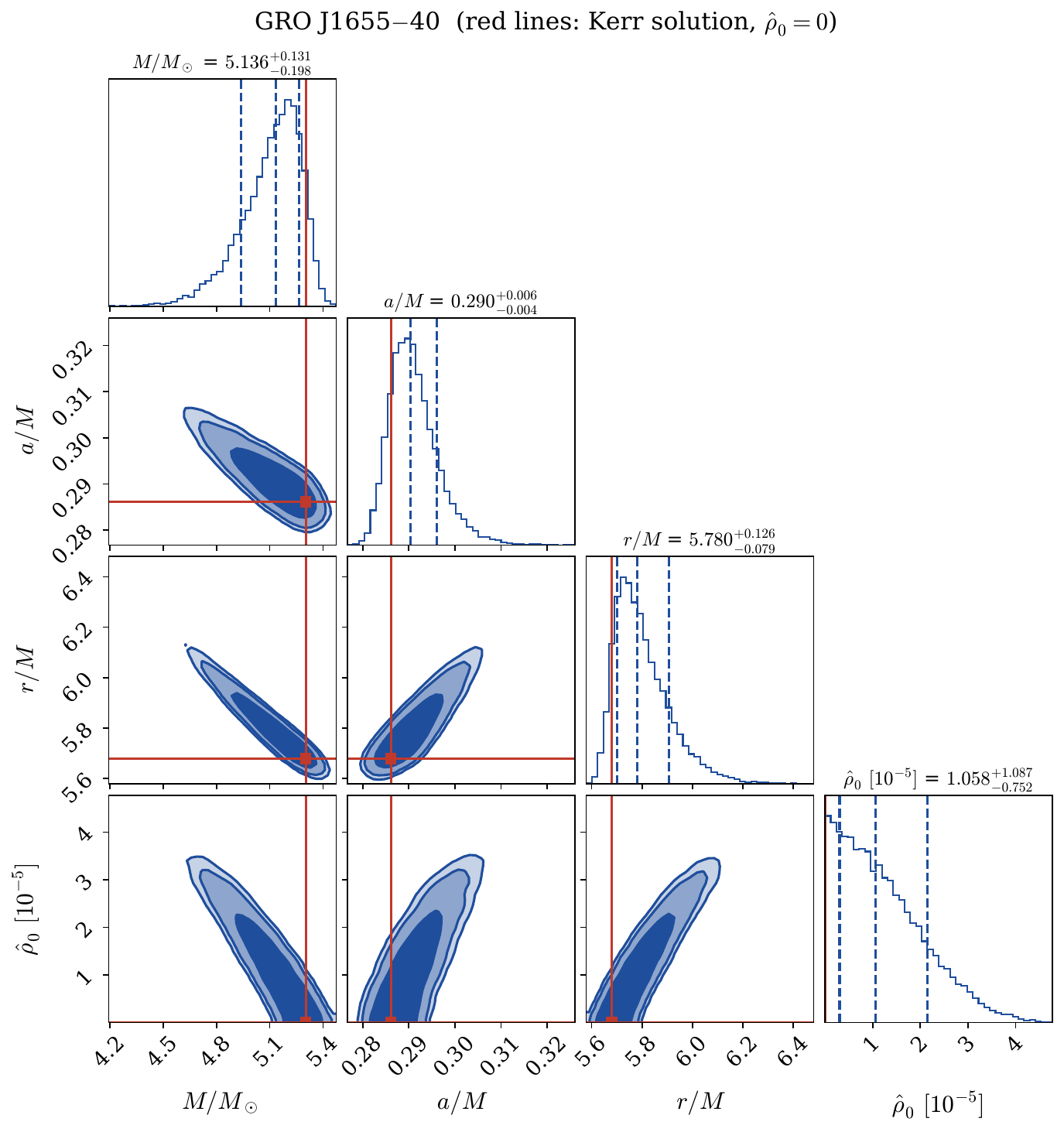}
\caption{Posterior distributions of the parameters of the rotating black hole in the exponential halo ($\hat r_0=20$) from the RP-model fit to the QPOs of GRO~J1655$-$40. Contours: 68\%, 90\% and 95\% credible regions. Red lines: Kerr solution.}
\label{fig:cornerGRO}
\end{figure*}

\begin{figure*}[p]
\centering
\includegraphics[width=0.85\textwidth]{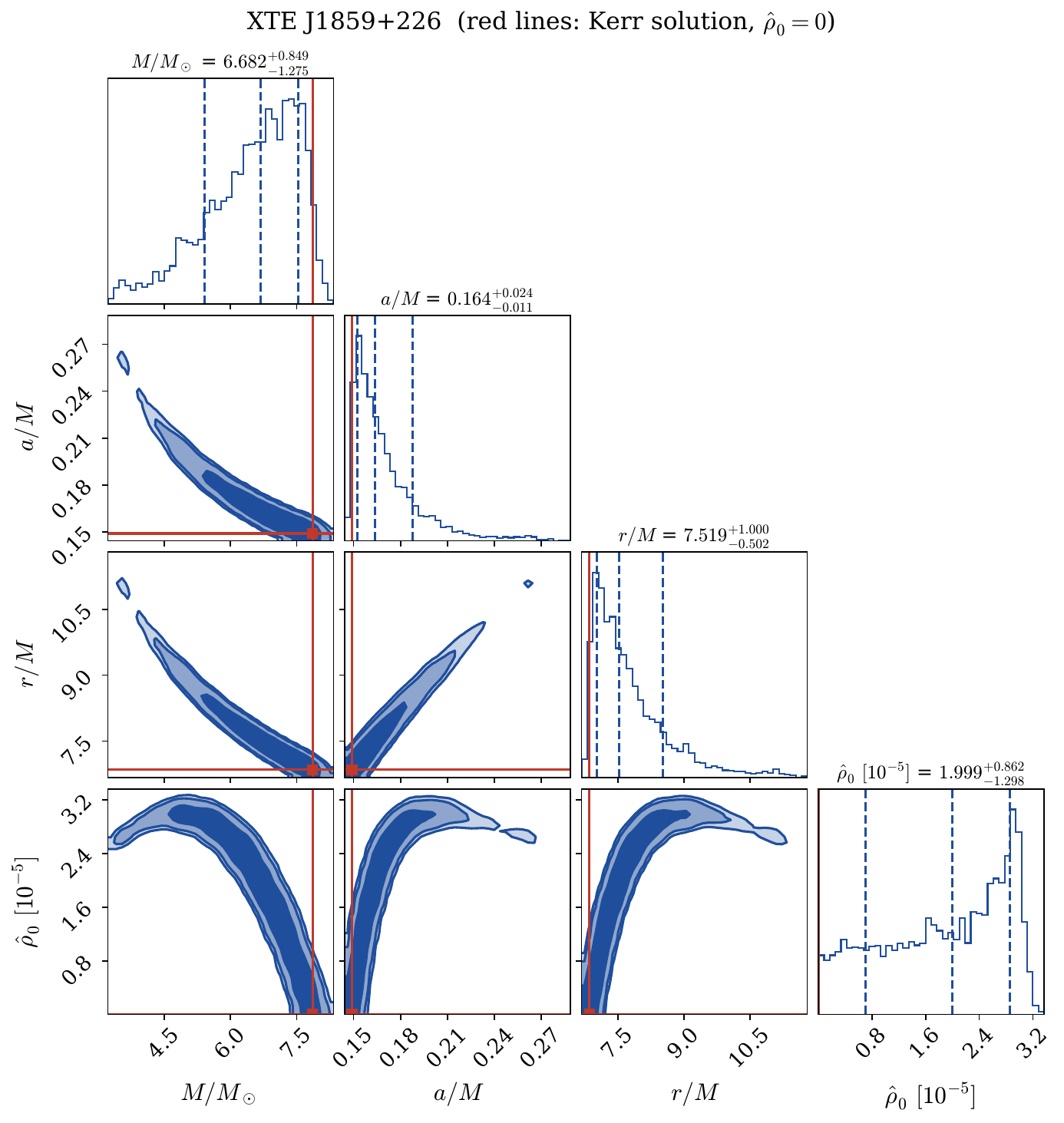}
\caption{Same as Fig.~\ref{fig:cornerGRO} for XTE~J1859+226. The posterior of $\hrho$ rises gently up to $\hrho\simeq3.0\times10^{-5}$, beyond which no stable circular orbit reproduces the observed triplet exactly, and falls off sharply above it.}
\label{fig:cornerXTE}
\end{figure*}

Figures~\ref{fig:cornerGRO} and \ref{fig:cornerXTE} show the one- and two-dimensional marginal posterior distributions of the RP-model fits of Sect.~\ref{sec:haloresults}, with the 68\%, 90\% and 95\% credible contours. Dashed lines in the one-dimensional panels mark the 16th, 50th and 84th percentiles, and the red lines mark the Kerr solution of Table~\ref{tab:kerr} ($\hrho=0$). The elongated contours in the $M$--$\hrho$, $a$--$\hrho$ and $r_e$--$\hrho$ planes trace the one-parameter family of exact fits discussed in Sect.~\ref{sec:haloresults}.

\bibliographystyle{spphys}
\bibliography{0refs}

\end{document}